\pdfoutput=1
\documentclass[11pt,two-sided]{article}
\usepackage[mathscr]{eucal}
\usepackage{enumerate}
\usepackage{epsfig,amsfonts}
\usepackage{pifont}
\usepackage{amsmath}
\usepackage{amsthm,amssymb}
\usepackage{bm,bbm}
\usepackage{graphicx}
\usepackage{relsize}
\usepackage{hhline}
\usepackage{cite}
\usepackage{psfrag}
\usepackage{mathrsfs} 
\usepackage{hyperref}
\usepackage{bm}
\usepackage{array}
\usepackage{tikz}
\usetikzlibrary{arrows}
\usetikzlibrary{arrows.meta}
\usetikzlibrary{calc}
\usetikzlibrary{fadings}
\usepackage{textcomp}
\usepackage{hyperref}
\usepackage{mathrsfs} 
\usepackage{simplewick}
\usepackage{lettrine}
\usepackage{tcolorbox}
\tcbuselibrary{breakable}
\usepackage[abspath]{currfile}

\DeclareMathAlphabet{\mathpzc}{OT1}{pzc}{m}{it}
\DeclareMathAlphabet\mathbfcal{OMS}{cmsy}{b}{n} 
\makeatletter
\@addtoreset{equation}{section}
\makeatother
\newcommand*\circled[1]{\tikz[baseline=(char.base)]{
    \node[shape=circle, draw, inner sep=1pt, 
        minimum height=8pt] (char) {#1};}}

\newcounter{app}
\newcounter{sapp}[app]
\def\theapp{\Alph{app}}
\newcommand{\app}[1]{
\refstepcounter{app}{\vspace{7mm}
\noindent\Large\bf Appendix
\theapp.
 \ #1 \par \vspace{5mm}}
\setcounter{equation}{0}
\def\theequation{\Alph{app}.\arabic{equation}}}

\renewcommand{\theequation}{\thesection.\arabic{equation}}

\usepackage{cancel}
\newcommand{\ds}{\displaystyle}

\def\bea{\begin{eqnarray}}
\def\eea{\end{eqnarray}}
\def\be{\begin{equation}}
\def\ee{\end{equation}}
\def\beq{\begin{equation}}
\def\eeq{\end{equation}}

\def\zsite{\kappa_s}
\def\zedge{\kappa_e}
\newcommand{\Si}{\textrm{\scalebox{1.2}{$\sigma$}}}
\newcommand{\Om}{\textrm{\scalebox{1.5}{$\omega$}}}

\def\iq{{\xi}}
\def\ii{{\ri}}
\def\EXP{{\re}}

\def\be{\begin{equation}}
\def\ee{\end{equation}}
\def\bdm{\begin{displaymath}}
\def\edm{\end{displaymath}}
\def\bea{\begin{eqnarray}}
\def\eea{\end{eqnarray}}

\def\ri{{\rm i}}
\def\half{{\textstyle\frac{1}{2}}}

\def\XXint#1#2#3{{\setbox0=\hbox{$#1{#2#3}{\int}$}
    \vcenter{\hbox{$#2#3$}}\kern-.5\wd0}}

\def\qs{\mathsf q}

\newcommand{\rd}{\mbox{d}}
\newcommand{\re}{\mbox{e}}

\newcommand{\Dfr}{{\mathfrak D}}

\def\U{\mathbf U}
\def\V{\mathbf V}
\def\V{\boldsymbol{V}}
\def\U{\boldsymbol{U}}
\def\iq{{\xi}}
\def\Lc{{\mathcal L}}

\def\Rc{{\mathcal R}}

\def\as{{\mathsf a}}
\def\bs{{\mathsf b}}
\def\cs{{\mathsf c}}
\def\dds{{\mathsf d}}

\def\dla{\langle{\kern-.1em} \langle}
\def\dra{\rangle{\kern-.1em} \rangle}

\usepackage[font=small,labelfont=bf,format=hang]{caption}

\begin{document}

\begin{titlepage}
\vglue 2 cm

\begin{center}
\begin{LARGE}

{\bf A distant descendant of the six-vertex model
}
\end{LARGE}

\vspace{1.3cm}
\begin{large}

{\bf Vladimir V. Bazhanov$^{1}$ and Sergey M. Sergeev$^{1,2}$}

\end{large}

\vspace{1.cm}
$^1$ Department of Fundamental and Theoretical Physics,
         Research School of Physics\\
    Australian National University, Canberra, ACT 2601, Australia\\\ \\
$^2$ Faculty of Science and Technology,
      University of Canberra,
 Bruce, ACT 2617,\ \ \ \ \ \ \ \\ Australia\\\ \\

\vspace{.2cm}

\vspace{1cm}
\end{center}

\begin{center}

\parbox{14.9cm}{%
{\bf Abstract}. 
In this paper we present a new solution of the star-triangle relation
having positive Boltzmann weights.  The solution defines an exactly
solvable two-dimensional Ising-type (edge interaction) model of
statistical mechanics where the local ``spin variables'' can take arbitrary
integer values, i.e., the number of possible spin states at each site of the lattice is infinite.
 There is also an equivalent ``dual'' formulation of
the model, where the spins take continuous real values on the
circle. From algebraic point of view this model is closely related
to the to the 6-vertex model. It is connected with the
construction of an intertwiner for two infinite-dimensional
representations of the quantum affine algebra $U_q(\widehat{sl}(2))$
without the highest and lowest weights.  
The partition function of the model in the large lattice limit is
calculated by the inversion relation method. Amazingly, it coincides with 
the partition function of the off-critical 8-vertex free-fermion
model. 
}
\end{center}

\vskip2cm

\vfill

\end{titlepage}
\setcounter{page}{2}

\setcounter{tocdepth}{2}
\newpage
\section{Introduction}
There is an important class of
 integrable two-dimensional lattice models of statistical
mechanics \cite{Baxter:1982zz} 
with only a pair interaction between neighbouring spins,
i.e., where two spins interact only if they are connected by an edge
of the lattice.  We will call them as ``edge interaction'' or
``Ising-type'' models. The commutativity of  
transfer matrices for these models can usually be derived from 
the star–triangle relation \cite{Onsager:1944} 
which is a special form of the Yang-Baxter equation
\cite{McGuire:1964,Yang:1967,Baxter:1972}.

Over the past forty years a number of such models 
was discovered. 
The most notable discrete-spin models
in this class include the Kashiwara-Miwa \cite{Kashiwara:1986}
and chiral Potts \cite{vG85,Au-Yang:1987syu,Baxter:1987eq} models (both of
them also contain the Ising model \cite{Onsager:1944} and Fateev-Zamolodchikov
$Z_N$-model \cite{FZ82} as particular cases) see \cite{Bax02rip} for
a review.  
There are also important continuous spin models, including
Zamolodchikov's ``fishing-net'' model \cite{Zam-fish}, which describes
certain planar Feynman diagrams in quantum field theory, and the
Faddeev-Volkov model \cite{FV95}, connected with quantization
\cite{BMS07a} of discrete conformal transformations \cite{BSp,
  Steph:2005}. 
%
Quite interestingly, 
all solutions of the star-triangle relation associated with
these models (both with discrete and
continuous spin variables) can be obtained 
as special cases of 
a rather general ``master solution'' 
of this relation, which was found in
\cite{Bazhanov:2010kz}.\footnote{To be more    
  precise, the ``master solution'' only contains 
the solutions, which have a single
  one-dimensional spin at each lattice site. For this reason, it
  cannot contain the $D\ge2$ fishing-net model which has
  multi-dimensional spins.}\ %
Algebraically, the master solution is related to the modular double 
\cite{Faddeev:1999,Spiridonov-essays} of
the elliptic Sklyanin algebra \cite{Skl82}
and Spiridonov's elliptic beta integral
\cite{Spiridonov-beta}.

In this paper we present yet another solution of the star-triangle relation,
with positive Boltzmann weights,  
which, apparently, cannot be obtained from the master solution and its
specializations (at least, to the authors' knowledge).
Our new solution
involves spin variables taking arbitrary {integer} values  
\,$-\infty< a,b<+\infty$,\,
and {reflection-symmetric} Boltzmann weights, which are unchanged upon
interchanging the edge spins $a,b$. The weights depend on the absolute value
of the spin difference,
\begin{equation}\label{W-def}
W_{x}(a-b)=W_{x}(b-a)=
x^{a-b}\;\frac{(-\qs^{a-b+1}\;x^2;\qs^2)_\infty}
{(-\qs^{a-b+1}/x^{2};\qs^2)_\infty}\;.
\eeq
and a {multiplicative} spectral variable $x$ (see how
it enters the star-triangle relation \eqref{str-main}
below). Here 
$(x;\qs^2)_\infty$ denotes the $\qs$-Pochhammer symbol 
\beq
(x;\qs^2)_\infty=\prod_{k=0}^\infty \,(1-x\,\qs^{2k})\,,
\eeq
involving a fixed parameter $\qs$, such that $|\qs|<1$. It is 
convenient to also define the ``crossing parameter'',
\beq\label{xidef}
\xi=\ri \sqrt{\qs}\,,\qquad \qs=-\xi^2\,.
\eeq

We state that the above weights satisfy the star-triangle relation of
the form,
\beq
\begin{array}{l}
\ds \sum_{d\in {\mathbb Z}}\;
{W}_{\xi/x}(a-d)\;W_{x\,y}(c-d)\;{W}_{\xi/y}(b-d)\,
\\[.5cm]
\phantom{W_{\xi/x}(a-d)\,W_{\xi/y}(b-d)\,}
={\cal R} \;W_x(b-c)\;{W}_{\xi/(x\,y)}(a-b)\;W_y(a-c)\,,
\end{array}\label{str-main}
\eeq
where ${\cal R}$ is an explicitly known scalar
factor 
\beq\label{R-def}
{\cal R}=\kappa_s\,\frac{\Si(x^2)\,\Si(y^2)}{\Si(x^2 y^2)}\,,\qquad\ 
\Si(x)=
\frac{(-\qs/x;\qs)_\infty}{(x;\qs)_\infty}\,,\qquad
\kappa_s=\frac{(\qs;\qs)_\infty}{(-\qs;\qs)_\infty}\,,
\eeq
which depends on the
spectral variables $x$ and $y$, but is independent of
the spins $a,b,c\in{\mathbb Z}$. It is important to note that if the parameter $\xi$ in \eqref{xidef} and the
spectral variables $x$ and $y$ 
are real and belong to the domain  
\beq\label{physdom}
0<\xi<1\,,\qquad \xi<x,y,xy<1\,,
\eeq
then all weights entering \eqref{str-main} are {\em real and positive}.
Note also, that 
for large $n$ the expression inside the sum 
in \eqref{str-main} behaves as $O(\qs^{|n|})$, 
so for $|\qs|<1$ the sum always converges as a geometric series. 

Let us briefly explain how the above results were obtained. The
Boltzmann weights \eqref{W-def} originate from the calculation 
of the $R$-matrix  
intertwining two particular infinite-dimensional representations of the quantum
affine algebra $U_\qs(\widehat{sl}(2))$. These representations do not have 
the highest and lowest weights and their Cartan elements are realized
as shift operators. A similar problem, but for the case of
cyclic representations of $U_\qs(\widehat{sl}(2))$, with $\qs$ being a root of
unity, has been previously considered in \cite{Bazhanov:1989nc}.
Following the results of that work one would expect
that the matrix elements of the interwining $R$-matrix in our case 
factorize into a product of four 
factors
depending on two spins only. Indeed, our analysis 
exactly confirms such factorization. In particular, the associated
two-spin weights 
$W_x(a-b)$ are determined by the following recurrence 
\beq\label{r2}
\frac{{W}_x(n)}{{W}_x(n-2)}\;=\;
\frac{x^2+\qs^{n-1}}{1+\qs^{n-1}\, x^2}\,,\qquad n\in{\mathbb Z}\,,
\eeq
and inversion
relations,
\beq\label{un2}
\sum_{n\in{\mathbb Z}}\,W_{\xi x}(a-n)\,
W_{\xi/x}(n-b)\simeq \delta_{a,b}\,,
\qquad a,b,n\in{\mathbb Z}\,,
\eeq
where $\xi$ is defined in \eqref{xidef} and 
the symbol $\simeq$ means the proportionality up to a scalar
factor independent of spins.
Notice, that the relation \eqref{r2}
is a {\em second order} recurrence
relation, which do not allow to determine
$W_x(n)$ without additional information. 
For instance, starting from the value of $W_x(0)$ one can
only calculate the values of $W_x(n)$ for even $n=\pm 2,\pm
4,\ldots$, however, in order to calculate them for odd
$n=\pm1,\pm3\ldots$, one needs to  
specify another initial value, say $W_x(1)$. Both
$W_x(0)$ 
and $W_x(1)$ are not just constants,
but functions of the spectral variable $x$,
and only one of them could be absorbed into the
overall normalization of $W_x(n)$. 

The above ``odd-even'' problem is, in fact, well known 
\cite{Au-Yang:2018} in connection
with the chiral Potts model (though in that case it is
slightly simpler, because, when $\qs$ is a root of unity, the values of spins
are identified {\em modulo} some integer $N$,\ $N\ge2$, and,
therefore, can 
only take a finite number of values $n\in {\mathbb Z}_N$).\footnote{%
Actually, the problem only arises for even values of $N$,
since for odd $N$ 
the second order recurrence \eqref{r2}
covers all elements of the set ${\mathbb Z}_N$ due to the cyclic symmetry.}
The resolution of this problem, suggested in
\cite{BBP90}, essentially reduces to an exclusion of odd numbers from
all consideration.  Technically this is achieved by the replacement $n\to2n$
which transforms \eqref{r2} into a first order recurrence for $W$. 
Unfortunately, the above recipe does not work in our $|\qs|<1$ 
case, since the 
weights $W_x(n)$, obtained in this way, do not satisfy the
inversion relation \eqref{un2} (despite satisfying it in the root of
unity case). 

In our approach we retain the second order recurrence relation in the
original form \eqref{r2} and substitute its most general solution,
including two unknown functions $W_x(0)$ and $W_x(1)$, into the
inversion relation \eqref{un2}.  Remarkably, this relation allows one 
to uniquely fix the weights $W_x(n)$ (to within an overall
normalization and some trivial equivalence transformation
factors). The result is given in \eqref{W-def}. The handling of
the inversion relation \eqref{un2} is based on some important extensions of the
Ramanujan ${}_1\psi_1$ bilateral summation formula \cite{Vildanov:2012}.

Initially, we have obtained the star-triangle relation
\eqref{str-main} by the perturbation theory around the point $xy=1$ (where
\eqref{str-main} reduces to the inversion relation \eqref{un2}) and
then thoroughly verified it by numerical calculations. Subsequently,
but rather accidentally, we have realized that the star-triangle
relation \eqref{str-main} could be deduced from the {\em ``constant beta
pentagon equation for the circle locally compact Abelian group
${\mathbb T}$''},\  by Garoufalidis and Kashaev \cite{Garoufalidis:2017xah}. 
Evidently, the 
connection is far from being obvious but, certainly, worth to be
explored further.
  
As is well known \cite{Baxter:1982zz} every solution of the
star-triangle relation can be used to define exactly solvable edge
interaction models on various two-dimensional lattices. For purposes of this
introduction it is enough to consider an homogeneous square lattice.
In this case the partition function reads
\beq
{Z}=\sum_{a_{m}\in {\mathbb Z}}\ 
\ \prod_{(ij)}\big(\,\rho(x)\,W_{x}(a_i-a_j)\,\big)\ \ 
\prod_{(kl)}\big(\,\rho(\xi/x)\,W_{\xi/x}(a_k-a_l)\,\big)\ 
\label{z-main}
\end{equation}
where the first product is taken over all horizontal edges $(ij)$, the
second over all vertical edges $(kl)$ and the sum is taken over the
spins $\{a_m\}$ on interior sites of the lattice.  The boundary spins
are assumed to be fixed. In writing \eqref{z-main} we have included
arbitrary normalization factors $\rho(x)$ and $\rho(\xi/x)$, which might be
imposed by the physical interpretation of the model. 
However, to keep the formulae more readable we prefer to drop these
factors for the most of our considerations, only restoring them
when necessary.

Note that, since the edge weights in \eqref{z-main} depend on 
spin differences, there exist an equivalent ``dual'' formulation of
the model.  It is also a square lattice edge-interaction model, 
where the spins take
continuous real values on the circle $0\le \phi_i,\phi_j<2\pi$. The
correponding weights \ $\overline{\Om}_{\xi/x}(\phi_i-\phi_j)$,
$\overline{\Om}_{x}(\phi_k-\phi_l)$ for the horizontal and vertical edges are real and positive for $0<\xi<x<1$. 
They are defined by the Fourier transform,
\begin{equation}\label{Fourier0}
\overline{\Om}_{x}(\varphi)\;=\;{\kappa_s}^{-1}\,{\Si(x^2)}\, 
\sum_{n\in{\mathbb Z}} \, {W}_{x}(n)
\, \re^{\ri\varphi n}=\Si(x\re^{\ri \phi})\,\Si(x\re^{-\ri \phi})\;,
\end{equation}
where $\kappa_s$ and $\Si(x)$ are given in \eqref{R-def}. 
In the dual formulation the star-triangle relation \eqref{str-main}
becomes a concise integral identity 
\beq \label{beta-int}
\frac{1}{2\pi  \ri}\oint\, \frac{d w}{w}\,\prod_{j=1}^3
\frac{\Si(u_j\, w)}{\Si(v_j\, w)}=
\kappa_s\,\prod_{j=1}^3\prod_{k=1}^3 {\Si(v_j/u_k)}^{-1}\,,\qquad 
\frac{v_1\, v_2\, v_3}{u_1\, u_2\, u_3}=\qs^2\,,
\eeq
where the integration countor is such that 
\beq\label{beta-contour}
\max(|\qs/v_j|)<|w|<\min(|1/u_j|)\,.
\eeq
Here $\{u_j\}$ and $\{v_j\}$ are complex numbers, 
satisfying the constraints shown above, but otherwise arbitrary.
Recalling that each function $\Si(x)$ is a ratio of two $\qs$-Pochhammer
symbols, the numerator and denominator of the integrand in
\eqref{beta-int} each contains products of six
$\qs$-Pochhammer symbols.

Using the inversion relation method \cite{Str79, Zam79,Bax82inv} 
we calculate the partition function \eqref{z-main} of the model for
the main physical regime\  $0<\xi<x<1$\  
in a large lattice limit. Surprizinly, it coincides with the partition
function of the off-critical 8-vertex model \cite{Baxter:1972} at the
free-fermion point.  

The organization of the paper is as follows. In Sect.2 we consider the
problem of constructing of $R$-matrices, intertwining some particular
infinite-dimensional representations of the quantun affine algebra
$U_{\qs}(\widehat{sl}(2))$ and derive a set of equation defining such
$R$-matrices. 
In Sect.3 we solve these equations using the ideas of factorized
R-matrices and the ``face-vertex'' correspondence. In Sect.4 we review
the theory of solvable lattice models on arbitrary planar graphs. In
Sect.5 we calculate the partition function of the model in the
thermodynamic limit. In Conclusion we discuss the main results and
indicate some directions for a future research.

\section{Yang-Baxter equations} 

We start with 
the standard $R$-matrix of the 6-vertex model,
\be
\begin{array}{l}
\Rc^{\rm (6v)}(\mu)_{++}^{++}\;=\;\Rc^{\rm (6v)}(\mu)_{--}^{--}\;=\;
\mu\, \qs-\mu^{-1} \qs^{-1}\,,\qquad\qquad \\[.6cm]
\Rc^{\rm (6v)}(\mu)_{+-}^{-+}\;=\;\Rc^{\rm (6v)}(\mu)_{-+}^{+-}\;=\;
\qs-\qs^{-1}\,,\\[.6cm]
\Rc^{\rm (6v)}(\mu)_{+-}^{+-}\;=\;\Rc^{\rm (6v)}(\mu)_{-+}^{-+}\;=\;
\mu -\mu^{-1}\,,
\end{array}\label{R6v}
\ee
and the Yang-Baxter equation (YBE) defining the ${\bf L}$-operator 
\beq\label{RLL}
\sum_{j_1,j_2=\pm1}\;{\bf L}(\mu_1)_{i_1}^{j_1}\;{\bf
  L}(\mu_2)_{i_2}^{j_2}\ 
\Rc^{\rm (6v)}(\mu_2/\mu_1)_{j_1,j_2}^{k_1,k_2}
=
\sum_{j_1,j_2=\pm1}\;
\Rc^{\rm (6v)}(\mu_2/\mu_1)_{i_1,i_2}^{j_1,j_2}\ 
{\bf L}(\mu_2)_{j_2}^{k_2}\;{\bf L}(\mu_1)_{j_1}^{k_1}\;,
\eeq
where the matrix indices $i_1,i_2,k_1,k_2$,\  
take two values $\pm1$ and $\qs$ is a fixed
parameter, 
such that $|\qs|<1$. 
The operator ${\bf
  L}(\mu)$ is a two-by-two matrix whose elements are operators acting
in the ``quantum space'' and $\mu$ is
the ``spectral variable''. Below we will often use the ``crossing
parameter'', defined as
\beq
\xi=\ri\sqrt{\qs}\,,\qquad \qs=-\xi^2\,.
\eeq
For the simplest ${\bf L}$-operator,
which is a first order Laurent polynomial in the spectral variable $\mu$,  
\beq\label{Lop2}
{\mathbf L}(\mu)=\begin{pmatrix}
\begin{array}{cc}
\mu\, K-
\mu^{-1}\, K^{-1} &
F\\[.4cm]
E&
\mu \,K^{-1}-\mu^{-1}\,K
\end{array}
\end{pmatrix}
\eeq
the YBE \eqref{RLL} just reduces to the defining relations of the
quantum universal enveloping algebra $U_{\qs}(sl(2))$,  
\beq
K\,E=\qs\,E\,K,\qquad K\,F=\qs^{-1}\,E\,K\,,\qquad
[E,F]={(\qs-\qs^{-1})}\,(K^2-K^{-2})\,.\label{uqksl2a}
\eeq
Thus any of its representations 
leads to a solution of \eqref{RLL}.
Consider a particular infinite-dimensional representation $\pi_s$ 
of this algebra, 
depending on the parameter $s$,  
\beq\label{pis2}
\pi_s:\qquad 
\pi_s[K]=\V\,,\qquad
\pi_s[E]=\U^{-1}\,\big(s \,\V^{-1}-s^{-1}\,\V\big)\,,\qquad
\pi_s[F]=\U\,\big(s\, \V-s^{-1}\,\V^{-1}\big)\,,
\eeq
defined by the action of the operators $\U$ and $\V$ 
on the infinite set of basis vectors
$|a\rangle$, $a\in{\mathbb Z}$,
\beq
\U\,\V=\qs\,\V\,\U\,,\qquad (\U)_{a,a'}=\qs^a\,\delta_{a,a'}\,,\qquad
(\V)_{a,a'}=\delta_{a,a'+1}\,,\qquad a,a'\in{\mathbb Z}\,.\label{weyl2}
\eeq
Note, that the Cartan element $\pi_s[K]=\V$ for this representation
is realized as the shift operator.  
The ${\bf L}$-operator \eqref{Lop2} then takes the form
\beq
{\mathbf L}(\mu\,|\,s)=\pi_s\big[{\mathbf L}(\mu)\big]=
\begin{pmatrix}
\begin{array}{cc}
\mu\,\V-
\mu^{-1}\, \V^{-1} &
\ \U\,\big(s\, \V-s^{-1}\,\V^{-1}\big)
\\[.4cm]
\U^{-1}\,\big(s \,\V^{-1}-s^{-1}\,\V\big)
&
\mu\,\V^{-1}-
\mu^{-1}\, \V
\end{array}
\end{pmatrix}\label{Lmus}
\eeq

Now take
two such ${\bf L}$-operators 
\beq
{\mathbf L}_1= {\mathbf L}_1(\mu_1\,|\,s_1)=\pi_{s_1}[{\bf L}(\mu_1)]\,,\qquad
{\mathbf L}_2={\mathbf L}_2(\mu_2\,|\,s_2)=\pi_{s_2}[{\bf L}(\mu_2)]\,.
\eeq
associated with two different representations $\pi_{s_1}$ and
$\pi_{s_2}$, labelled by the subscripts $1$ and $2$. 
Consider the problem of construction of an
intertwining operator 
\beq
{\bf S}_{12}
={\mathbf
  S}_{12}(\mu_1/\mu_2\,|\,s_1\,,s_2)\,, 
\eeq
which acts in the tensor product $\pi_{s_1}\otimes \pi_{s_2}$
and satisfies the YBE
\beq\label{Sybe2}
\begin{array}{r}
\Big({\mathbf L}_1(\mu_1\,|\,s_1)\circ{\mathbf L}_2(\mu_2\,|\,s_2)\Big)\,
{\mathbf
  S}_{12}(\mu_1/\mu_2\,|\,s_1\,,s_2)=
\phantom{\qquad\qquad\qquad\qquad\qquad\qquad}  
\\[.3cm] 
={\mathbf S}_{12}(\mu_1/\mu_2\,|\,s_1\,,s_2)\,
\Big({\mathbf L}_2(\mu_2\,|\,s_2)\circ{\mathbf L}_1(\mu_1\,|\,s_1)\Big)\,.
\end{array}
\eeq
Here the ``{circle-product}'' 
notation $\big({\bf L}_1\circ{\bf L}_2\big)$ denotes the matrix
product in the two-dimensional space and the tensor product in  
the infinite-dimensional quantum space $\pi_{s_1}\otimes \pi_{s_2}$. 
It is useful to view \eqref{Sybe2} 
as defining relations of the Yang-Baxter
algebra realizing permutations of factors in the 
circle-products of ${\bf L}$-operators, by similarity transformations in the
quantum space. 
Then, as is well known,  
the associativity condition for this algebra reduces to the YBE for
intertwining operator ${\bf S}$,     
\beq\label{SSS}
{\mathbf S}_{12}\,
{\mathbf S}_{13}\,
{\mathbf S}_{23}
={\mathbf S}_{23}\,
{\mathbf S}_{13}\,
{\mathbf S}_{12}\,,
\eeq
where for brevity we have omitted 
(rather obvious) arguments of 
${\mathbf S}_{ij}$. 

Clearly, the above ``permutation theory''  
interpretation of \eqref{Sybe2} 
requires some additional consistency
relations, 
which we call the inversion relations.
The first of them reads 
\beq\label{unit1}
{\mathbf S}_{12}(\mu_1/\mu_2\,|\,s_1\,,s_2)
\,{\mathbf S}_{21}(\mu_2/\mu_1\,|\,s_2\,,s_1)\simeq {\bf I}
\eeq 
where the RHS is proportional to the unit operator in
$\pi_{s_1}\otimes \pi_{s_2}$ to within a scalar
factor. The above relation is a simple corolary of fact that a permutation
followed by the inverse permutation 
\beq
\big({\bf L}_1\circ\,{\bf L}_2\big)\to
\big({\bf L}_2\circ\,{\bf L}_1\big)\to \big({\bf L}_1\circ\,{\bf
  L}_2\big)\,,
\eeq
should reduce to the identity
transformation. 
Indeed, using \eqref{Sybe2} twice, it is easy to see
  that the products  
$\big({\bf L}_1\circ\,{\bf L}_2\big)$ and 
$\big({\bf L}_2\circ\,{\bf L}_1\big)$ commute with the LHS of
\eqref{unit1}. Therefore, the latter must be proportional to the unit operator, 
otherwise the whole scheme becomes inconsistent. To obtain the second
relation, define the inverse ${\bf L}$-operator
\beq
{\bf L}_1(\mu_1\,|\,s_1)\;{\bf L}_1(\mu_1\,|\,s_1)^{-1}={\bf I}
\eeq
where the product in the LHS implies both the matrix product in
the two-dimensional space and
the operator product in the quantum space $\pi_{s_1}$. The symbol\  ${\bf
  I}$\  here denotes the unit operator in both spaces. Now, let us
multiply each side of \eqref{Sybe2} by ${\bf L}_1(\mu_1\,|\,s_1)^{-1}$
both from the left and from the right. Transposing the resulting
equation in the space $\pi_{s_1}$, one obtains
\beq\label{Sybe3}
\begin{array}{r}
\Big({\bf L}_2(\mu_2\,|\,s_2)\circ {\bf L}_1(\widetilde{\mu}_1\,|\,\widetilde{s}_1)\Big)\; 
{\bf S}_{12}(\mu_1/\mu_2\,|\,s_1\,,s_2)^{t_1}=
\phantom{\qquad\qquad\qquad\qquad\qquad\qquad}  
\\[.4cm] 
={\bf S}_{12}(\mu_1/\mu_2\,|\,s_1\,,s_2)^{t_1}\;
\Big({\bf L}_1(\widetilde{\mu}_1\,|\,\widetilde{s}_1)\circ {\bf L}_2(\mu_2\,|\,s_2)\Big)\; 
\end{array}
\eeq
where
\beq
\widetilde{\mu}_1=\mu_1\,\qs^{-1}\,,\qquad
\widetilde{s}_1=s_1^{-1}\,\qs^{-1}\,,
\eeq
where the superscript $t_1$ denotes the transposition in the space $\pi_{s_1}$.
Here we have used the relation
\beq
\big({\bf L}_1(\mu_1\,|\,s_1)^{-1}\big)^{t_1}=\frac{1}{[\mu_1
    s_1][\mu_1/(s_1 \qs)]}\,{\bf
  L}_1(\widetilde{\mu}_1\,|\,\widetilde{s}_1)\,,\qquad [x]=x-x^{-1}\,,
\eeq
which was obtained by a straitforward calculation from \eqref{Lop2},
\eqref{pis2} and \eqref{weyl2}.
Now taking \eqref{Sybe3} and repeating the arguments that have led to
\eqref{unit1} one obtains the second inversion relation
\beq\label{unit2}
{\mathbf S}_{12}(\mu_1/\mu_2\,|\,s_1\,,s_2)^{t_1}
\ {\mathbf S}_{21}\big(\qs^2\mu_2/\mu_1 \,|
-(s_2\qs)^{-1},-(s_1\qs)^{-1}\big)^{t_2}\simeq {\bf I}\,,
\eeq 
which will be used later on.

\bigskip
Below, it will be convenient to use a slightly different set of 
spectral variables. Namely for the ${\bf L}$-operator 
\beq
{\bf L}(\mu\,|\,s)=\bm{\Lc}(x,\,x',\,y)\,,\qquad \mu=\xi x x'/y^2\,,\qquad
s=x/(x'\xi)\,,\qquad \xi=\ri\sqrt{\qs}\,,
\eeq
we will use new spectral variables $x$, $x'$ and $y$.
Explicitly, using the formulae \eqref{Lop2}, \eqref{pis2} and 
\eqref{weyl2}, one obtains
\beq\label{Lop-new}
\big(\bm{\Lc}(x,\,x'/\xi,\,y)\big)_{a}^{a'}=
\begin{pmatrix}
\begin{array}{cc}\ds
\frac{x x'}{y^2}\,\delta_{a,a'+1}-
\frac{y^2}{x x'}\; \delta_{a,a'-1} &\ds
\ \qs^a\,
\Big(\,\frac{x}{x'}\, \delta_{a,a'+1}-\frac{x'}{x}\,\delta_{a,a'-1}\Big)
\\[.4cm]
\ds \qs^{-a}\,\Big(\frac{x}{x'} \,\delta_{a,a'-1}
-\frac{x'}{x}\,\delta_{a,a'+1}\Big)
&\ds
\frac{x x'}{y^2} \,\delta_{a,a'-1}-
\frac{y^2}{x x'}\; \delta_{a,a'+1}
\end{array}
\end{pmatrix}\,,
\eeq
where the indices $a,a'\in {\mathbb Z}$ refer to the quantum
space. Next, let us parameterize the matrix elements of the operator
${\mathbf S}_{12}(\mu_1/\mu_2\,|\,s_1\,,s_2)$
entering \eqref{Sybe2},
\beq\label{Scal}
{\mathbf
  S}_{12}
(\mu_1/\mu_2\,|\,s_1\,,s_2)
_{a\,,b}^{a',b'}=
\bm{{\cal S}}(x,x',y,y')_{a\,,b}^{a',b'}
\,,
\eeq
by a new set of spectral variables
\beq
\mu_1/\mu_2=x x'/(y y')\,,\quad
s_1=x/(x'\xi)\,,\quad s_2=y/(y'\xi)\,.
\eeq
Here the matrix indicies $a,a'$ refer to the space $\pi_{s_1}$ and the indices $b,b'$ to the space $\pi_{s_2}$.
With these definitions one can bring the YBE \eqref{Sybe2} to the form
\be
\begin{array}{r}
\ds\sum_{a',b'\in{\mathbb Z}}\ 
\bm{{\cal L}}(x,x',z)_{a}^{a'}\ 
\bm{{\cal L}}(y,y',z)_{b}^{b'}\ 
\bm{{\cal S}}(x,x',y,y')_{a'\,,b'}^{a'',b''}\phantom{\qquad
\qquad }
\\[.6cm]
\ds =\sum_{a',b'\in{\mathbb Z}}
\ \bm{{\cal S}}(x,x',y,y')_{a\;,b}^{a',b'}\ 
\bm{{\cal L}}(y,y',z)_{b'}^{b''}\ 
\bm{{\cal L}}(x,x',z)_{a'}^{a''}\,.
\end{array}\label{LLR2}
\ee
where $a,a'',b,b''\in{\mathbb Z}$.
Note, that there are no infinite summations: each sum above contains
exactly four terms, due to the 
special form \eqref{Lop-new} of the ${\bf L}$-operators. 

Finally, with the new notations the inversion relation \eqref{unit1}
and \eqref{unit2} read 
\beq\label{unit1a}
\sum_{a',b'\in{\mathbb Z}}\bm{{\cal S}}(x,x',y,y')_{a\,,b}^{a',b'}
\;\bm{{\cal S}}(y,y',x,x')_{b'\,,a'}^{b'',a''}\simeq \delta_{a,\,a''}
\,\delta_{b,\,b''}
\eeq
\beq\label{unit2a}
\sum_{a',b'\in{\mathbb Z}}\bm{{\cal S}}(x,x',y,y')_{a',b}^{a\;,b'}
\;\bm{{\cal S}}\big(\xi y',\,\xi y,\,x'/\xi,\,x/\xi\big)_{b'',a'}^{b'\,,a''}
\simeq \delta_{a,\,a''}
\,\delta_{b,\,b''}\,,
\eeq
where the symbol $\simeq$ means the proportionality up to a scalar
factor independent of spins.

\section{From vertex models to Ising-type models}
\subsection{Face-vertex correspondence\label{face-vert}}
In this section we solve the YBE \eqref{LLR2} together with the
inversion relations  
\eqref{unit1a}, \eqref{unit2a} and find an explicit expression 
for the operator $\bm{{\cal S}}(x,x',y,y')$ 
(the notation introduced in \eqref{Scal}).
The calculations are based on the idea of
factorized $R$-matrices, the Baxter's ``propagation
through the vertex'' techniques and the ``face-vertex'' correspondence
\cite{Baxter:1982zz}. Note, in particular, that  
this scheme was successfully used in \cite{BKMS}.

The ${\bf L}$-operator of the type \eqref{Lmus}
(but for cyclic representations of the Weyl algebra \eqref{weyl2} 
with $\qs$ being a root of unity) has previously appeared in
\cite{Bazhanov:1989nc} in the context of the chiral Potts
model \cite{vG85,Au-Yang:1987syu,Baxter:1987eq}.
The results of \cite{Bazhanov:1989nc} suggest that the matrix elements 
$\bm{{\cal S}}(x,x',y,y')_{a\;,b}^{a',b'}$, solving \eqref{LLR2},
factorize into a product of four factors depending on two spins only. 
Indeed, we will show that
\beq\label{Sfactor}
\bm{{\cal S}}(x,x',y,y')_{a\;,b}^{a',b'}=
 W_{x'/ y}(a-b)\; W_{x/y'}(b'-a')\;
\overline{W}_{x/y}(a-b')\;\overline{W}_{x'/y'}(b-a')
\eeq
where the functions $W_{x/y}(a-b)$ and $\overline{W}_{x/y}(a-b)$
depend on the ratio $x/y$ of two spectral variables and on the
difference $(a-b)$ of two spins.\footnote{Intially, one could assume that all
four functions in the RHS of \eqref{Sfactor} 
are different, but the a simple analysis (which we
skip here) shows that they
should pairwise coincide. The $R$-matrix factorization of the type
\eqref{Sfactor} usually take place when the
Cartan element $K$ of of the algebra \eqref{uqksl2a} 
is realized as the shift operator (see, e.g., \cite{BKMS} or Sect.5 of
ref.\cite{Bazhanov:2022wdj}).} 
It is well known, that the matrix elements $\bm{{\cal
    S}}(x,x',y,y')_{a\;,b}^{a',b'}$ can be used as Boltzmann weights
to define a vertex model on the square lattice. The factorization
\eqref{Sfactor} implies that this model can be reformulated as an
Ising-type model with only two-spin interaction through the edges of
the (medial) lattice.

Our considerations split into two steps. First we show that the
factorization \eqref{Sfactor} is a corollary of two rather simple 
(Yang-Baxter type) ``exchange relations'', involving only the two-spin
weights $W$ and $\overline{W}$ and some explicily known tri-spin
weights. Then we solve all these relations and 
explicitly determine the unknown
two-spin weights $W$ and $\overline{W}$.

The
proof of \eqref{Sfactor} is based on 
important factorization properties of the ${\bf L}$-operator
\eqref{Lop-new}, which we describe below.
Introduce an infinite set of two-dimensional 
vectors labelled by $a,a'\in{\mathbb Z}$ 
\be\label{Phi-def}
\begin{array}{rcl}
\ds\Phi(x,y)_{a,\sigma}^{a'}&=
&\ds\big(\ri \,\qs^{\frac{a}{2}}\, x/y\big)^{\sigma}\,\delta_{a,a'+1}
+\big(-\ri\, \qs^{\frac{a}{2}}\, y/x\big)^{\sigma}\,\delta_{a,a'-1}\,,\\[.4cm]
\ds\overline{\Phi}(x,y)_{a,\sigma}^{a'}&=&
\ds-\ri\,\big(\qs^{-\frac{a}{2}}\,x/y\big)^{\sigma}\,\delta_{a,a'+1}
-\ri\,\big(\qs^{-\frac{a}{2}}\, y/x\big)^{\sigma}\,\delta_{a,a'-1}\,,
\end{array}
\ee
with the components indexed by $\sigma=\pm1$. 
The half-integer powers of $\qs$ here and in some
of the equations below are understood as $\qs^{\frac{a}{2}}=(\sqrt{\qs})^a$,
where the sign of the square root should be chosen consistently
throughout all equation.
The above vectors satisfy important orthogonality relations
\begin{equation}\label{Phi-ort}
\sum_{a'\in{\mathbb Z}} \Phi(x,y)_{a,\sigma}^{a'}\,
\overline{\Phi}(x,y)_{a,\sigma'}^{a'}\;=\;[x^2/y^2]\,
\delta_{\sigma,\sigma'}\;,\qquad
\sum_{\sigma=\pm1} \overline{\Phi}(x,y)_{a,\sigma}^{a'}\,
\Phi(x,y)_{a,\sigma}^{a''}\;=\;[x^2/y^2]\,\delta_{a',a''}\;.
\end{equation}
where we used the notation $[x]=x-x^{-1}$.
It is convenient to also introduce another set of two-dimensional vectors 
\be\label{Omega-def}
\ds\Omega(x,y)_{a,\sigma}^{a'}=\Phi(x/\xi,y)_{a,\sigma}^{a'}\,,
\qquad
\ds\overline{\Omega}(x,y)_{a,\sigma}^{a'}=
\ds\overline{\Phi}(\xi x,y)_{a,\sigma}^{a'}\,,
\ee
which satisfy slightly different orthogonality conditions
\begin{equation}\label{omega-ort}
\sum_{a\in{\mathbb Z}} \Omega(x,y)_{a,\sigma}^{a'}\,
\overline{\Omega}(x,y)_{a,\sigma'}^{a'}\;=\;[x^2/y^2]\,
\delta_{\sigma,\sigma'}\;,\qquad
\sum_{\sigma=\pm1} \overline{\Omega}(x,y)_{a',\sigma}^{a}\,
\Omega(x,y)_{a'',\sigma}^{a}\;=\;[x^2/y^2]\,\delta_{a',a''}\;.
\end{equation}
where $\sigma,\sigma'=\pm1$. 
Using the explicit form of the two-by-two $L$-operator \eqref{Lop-new} 
it is easy to check that 
\begin{equation}\label{LL2}
\begin{tikzpicture}[baseline=(current  bounding  box.center)]
\draw [fill, opacity=0.3, blue,path fading=north] 
(1.5,0) -- (1.5,1.5) -- (2.1,1.7) -- (2.7,1.5) -- (2.7,0)  -- (1.5,0);
\draw [fill, opacity=0.3, blue,path fading=south] 
(1.5,-1.5) -- (1.5,0) -- (2.7,0) -- (2.7,-1.5) -- (2.1,-1.7) -- (1.5,-1.5);
\draw [-open triangle 45, thin] (1.5,-1.5) -- (1.5,1.5); \node [left] at (1.5,-1.5) {$x^{\phantom{'}}$};
\draw [-open triangle 45, thin] (2.7,-1.5) -- (2.7,1.5); \node [right] at (2.7,-1.5) {$x'$};
\draw [thick] (1.5,0) -- (0.8,0); 
\draw [-latex, thick] (0,0) -- (0.8,0);
\node [left] at (0,0) {$i$}; \node [right] at (4.2,0) {$i'$};
\node [below] at (0.45,0) {$z$};
\node [above] at (3.5,0) {$z$};
\draw [thick, dashed] (2.7,0) -- (1.5,0); 
\draw [thick] (4.2,0) -- (3.7,0); 
\draw [-latex,thick] (2.7,0) -- (3.7,0);

\node [anchor=south east] at (1.5,0) {$\Phi$};
\node [anchor=north west] at (2.7,0) {$\overline\Omega$};
\node [below] at (2.1,-.6) {$a$};
\node [above] at (2.1,.6) {$a'$};
\node [left] at (-0.6,0)
      {$\big(\bm{{\cal L}}(x,x',z)_{a}^{a'}\big)_{i,i'}
\;=\;\Phi(x,z)_{a,i}^{a'}\,
\overline{\Omega}(x',z)_{a,i'}^{a'}\;=\;$};
\end{tikzpicture}
\end{equation}
Note, that the indices $(i,i')=(+,+)$ refer to the top left element 
in \eqref{Lop-new}.  
Here we have used the following graphical notations 
\beq\label{thick-vert}
\begin{tikzpicture}[scale=.9,baseline=(current  bounding  box.center)] 
\draw [fill, opacity=.3,blue,path fading=north] 
(0,2) -- (1,1) -- (2,2) --(1,3);
\draw [fill, opacity=.3,blue,path fading=east] 
(1,1) -- (2,0) -- (3,1) --(2,2);
\draw [dashed, black, thick] (1,1) -- (2,2);
\draw [-{Triangle[width=4pt,length=10pt]},black, thick] (0,0) -- (.5,.5);
\draw [black, thick] (.4,.4) -- (1,1);
\draw [-open triangle 45, black, thin] (2,0) -- (0,2);
%
\node [left] at (0,0) {$i$};
\node [below] at (.5,.4) {$z$};
\node [above] at (0,2) {$x$};
\node at (1,2) {$a'$};
\node at (2,1) {$a$};
\node [right] at (-2.7,1) 
{$
\ds
{\Phi}(x,z)_{a,i}^{a'}\;=\ $};
\end{tikzpicture}
\qquad\qquad
\begin{tikzpicture}[scale=.9,baseline=(current  bounding  box.center)] 
\draw [fill, opacity=.3,blue,path fading=north] 
(0,2) -- (1,1) -- (2,2) --(1,3);
\draw [fill, opacity=.3,blue,path fading=west] 
(-1,1) -- (0,0) -- (1,1) --(0,2);
\draw [dashed, black, thick] (1,1) -- (0,2);
\draw [-{Triangle[width=4pt,length=10pt]},black, thick] (2,0) -- (1.5,.5);
\draw [black, thick] (1.6,.4) -- (1,1);
\draw [-open triangle 45, black, thin] (0,0) -- (2,2);
%
\node [right] at (2,0) {$i$};
\node [below] at (1.5,.4) {$z$};
\node [above] at (2,2) {$x$};
\node at (1,2) {$a$};
\node at (0,1) {$a'$};
\node [right] at (-3.7,1) 
{$
\ds
\Omega(x,z)_{a,i}^{a'}\;=\ $};
\end{tikzpicture}
\eeq
and
\noindent\beq\label{thick-vert-2}
\begin{tikzpicture}[scale=.9,baseline=(current  bounding  box.center)] 
\draw [fill, opacity=.3,blue,path fading=east] 
(1,1) -- (2,0) -- (3,1) --(2,2);
\draw [fill, opacity=.3,blue,path fading=south] 
(0,0) -- (1,1) -- (2,0) --(1,-1);
\draw [-open triangle 45, black, thin] (0,0) -- (2,2);
\draw [black, dashed,thick] (2,0) -- (1,1);
\draw [-{Triangle[width=4pt,length=10pt]},black, thick] (1,1) -- (.5,1.5);
\draw [black,thick] (.6,1.4) -- (0,2);
%
\node [above] at (.1,2) {$i$};
\node [below] at (.5,1.4) {$z$};
\node [above] at (2,2) {$x$};
\node  at (1,0) {$a'$};
\node  at (2,1) {$a$};
\node [right] at (-2.7,1)
{$
\ds
\overline{\Phi}(x,z)_{a,i}^{a'}\;=\ $};
\end{tikzpicture}
\qquad\qquad
\begin{tikzpicture}[scale=.9,baseline=(current  bounding  box.center)] 
\draw [fill, opacity=.3,blue,path fading=west] 
(-1,1) -- (0,0) -- (1,1) --(0,2);
\draw [fill, opacity=.3,blue,path fading=south] 
(0,0) -- (1,1) -- (2,0) --(1,-1);
\draw [-open triangle 45, black, thin] (2,0) -- (0,2);
\draw [black, dashed,thick] (0,0) -- (1,1);
\draw [-{Triangle[width=4pt,length=10pt]},black, thick] (1,1) -- (1.5,1.5);
\draw [black,thick] (1.4,1.4) -- (2,2);
%
\node [above] at (.1,2) {$x$};
\node [below] at (1.5,1.4) {$z$};
\node [above] at (2,2) {$i$};
\node [right] at (2,0) {$\phantom{i}$};
\node  at (1,0) {$a$};
\node  at (0,1) {$a'$};
\node [right] at (-3.7,1)
{$
\ds
\overline{\Omega}(x,z)_{a,i}^{a'}\;=\ $};
\end{tikzpicture}
\eeq

\noindent
There are two types of spin variables in these pictures. 
The variables $a,a'\in{\mathbb Z}$ are the ``face spins'' 
assigned to the shaded faces, while $i,i'=\pm1$ are the edge
spins (same as in the 6-vertex models) assigned to the directed solid lines in
the unshaded areas.  Moreover these lines (as well as their dashed
continuations into shaded areas) carry a spectral variable $z$. Another
spectral parameter $x$ is assigned to the directed thin lines,
which separate the shaded and unshaded areas.

After these preparations, let us now substitute \eqref{Sfactor} and
\eqref{LL2} into the YBE \eqref{LLR2}. The latter then reduces to two
rather simple exchange relations involving the two-spin weights  
$W_x(a-b)$ and $\overline{W}_x(a-b)$ 
and the ``face-vertex'' vectors $\Phi$ and
$\overline{\Omega}$, defined in \eqref{Phi-def} and \eqref{Omega-def},
which can be viewed as tri-spin weights. 
The first of these relations reads
\begin{equation}\label{psieq1}
W_{x/y}(a'-b')\;
\sum_{i} \overline{\Omega}(x,z)_{a,i}^{a'} \,\Phi(y,z)_{b,i}^{b'} \;
 \;=\;
W_{x/y}(a-b) \; \sum_{i} \overline{\Omega}(y,z)_{a,i}^{a'}\, \Phi(x,z)_{b,i}^{b'}\;,
\end{equation}
where $a,a',b,b'\in{\mathbb Z}$ (note, there is no summation over
these indices). 
Introduce the following graphical notations for the two-spin
weights,
\beq\label{W-shade}
(i):\quad 
\begin{tikzpicture}[scale=.9,baseline={(0,.8)}]
\draw [fill, opacity=.3,blue,path fading=east] 
(1,1) -- (2,0) -- (3,1) --(2,2);
\draw [fill, opacity=.3,blue,path fading=west] 
(0,0) -- (1,1) -- (0,2) --(-1,1);
\draw [-open triangle 45, black, thin] (0,0) -- (2,2);
\draw [-open triangle 45, black, thin] (2,0) -- (0,2);
%
\node [above] at (0,2) {$y$};
\node [above] at (2,2) {$x$};
\node  at (0,1) {$a$};
\node  at (2,1) {$b$};
\node [right] at (3.1,1) {$
\ds =\;{W_{x/y}(a-b)}\;,
$};
\end{tikzpicture}
\qquad
(ii):\quad
\begin{tikzpicture}[scale=.9,baseline=(current  bounding  box.center)] 
\draw [fill, opacity=.3,blue,path fading=north] 
(0,2) -- (1,1) -- (2,2) --(1,3);
\draw [fill, opacity=.3,blue,path fading=south] 
(0,0) -- (1,-1) -- (2,0) --(1,1);
\draw [-open triangle 45, black, thin] (0,0) -- (2,2);
\draw [-open triangle 45, black, thin] (2,0) -- (0,2);
%
\node [above] at (2,2) {$x$};
\node [above] at (0,2) {$y$};
\node at (1,2) {$b$};
\node at (1,0) {$a$};
\node [right] at (2.1,1) {$
\ds =\;\overline{W}_{x/y}(a-b)\;,
$};
\end{tikzpicture}
\eeq
by associating them with intersections of thin lines separating the
shaded and unshaded areas. Note that there are two types of such
intersections distinguished by the oriendation of the thin lines 
with respect to the shaded faces.
With these notations one can represent \eqref{psieq1} as
%
\begin{equation}\label{peq1}
\begin{tikzpicture}[scale=0.65
,baseline=20]
\draw [fill, opacity=0.3, blue,path fading=west] (-4,-2) -- (-2,-2) -- (0,2) -- (-0.5,3) -- (-2,3) -- (-4,-2);
\draw [fill, opacity=0.3, blue,path fading=east] (4,-2) -- (2,-2) -- (0,2) -- (0.5,3) -- (2,3) -- (4,-2);
\draw [-open triangle 45, black, thin] (-2,-2) -- (0.5,3) ; \node [below] at (-2,-2) {$x$};
\draw [-open triangle 45, black, thin] (2,-2) -- (-0.5,3); \node [below] at (2,-2) {$y$};
\draw [-latex, thick] (-1,0) -- (0,0); \draw [thick] (1,0) -- (-1,0);
\draw[dashed, thick] (-1,0) -- (-3.2,0);  \draw[dashed, thick] (1,0) -- (3.2,0); 
\node [above] at (-.4,0) {$z$};
\node [above] at (+.4,0) {$i$};
\node at (-1.35,2) {$a'$}; \node at (1.35,2) {$b'$};
\node [left] at (-2.1,-1) {$a$}; \node [right] at (2.1,-1) {$b$};
\node [anchor=north west] at (-1,0) {$\overline{\Omega}$};
\node [anchor=north east] at (1.1,-.1) {$\Phi$};
\node [right] at (4.5,0.5) {$=$};
\end{tikzpicture}
\hskip 10mm
\begin{tikzpicture}[scale=0.65,baseline=(current  bounding  box.center)]
\draw [fill, opacity=0.3, blue, path fading=west] (-2,-1) -- (-0.5,-1) -- (0,0) -- (-2,4) -- (-4,4) -- (-2,-1);
\draw [fill, opacity=0.3, blue, path fading=east] (2,-1) -- (0.5,-1) -- (0,0) -- (2,4) -- (4,4) -- (2,-1);
\draw [-open triangle 45, black, thin] (-0.5,-1)--(2,4)  ; \node [above] at (2,4) {$x$};
\draw [-open triangle 45, black, thin] (0.5,-1) -- (-2,4); \node [above] at (-2,4) {$y$};
\draw [-latex, thick] (-1,2) -- (0,2); \draw [thick] (1,2) -- (-1,2);
\draw[dashed, thick] (-1,2) -- (-3.2,2);  \draw[dashed, thick] (1,2) -- (3.2,2); \node [below] at (-.4,1.9) {$z$};
\node [below] at (.4,2) {$i$};
\node at (-1.35,0) {$a$}; \node at (1.35,0) {$b$};
\node [left] at (-2.1,3) {$a'$}; \node [right] at (2.1,3) {$b'$};
\node [anchor=south west] at (-1,2) {$\overline{\Omega}$};
\node [anchor=south east] at (1,2) {$\Phi$};
\end{tikzpicture}
\end{equation}

\noindent
where the summation is taken over the edge spins $i$ assigned to the
``internal'' (bounded) directed solid lines. 

\begin{figure}[ht]
\begin{center}
\begin{tikzpicture}[scale=0.8]
\draw [-open triangle 45, thin, black] (-4,-3) -- (1,2) ;
\draw [-open triangle 45, thin, black] (-3,-4) -- (2,1);
\draw [-open triangle 45, thin, black]  (3,-4) -- (-2,1);
\draw [-open triangle 45, thin, black] (4,-3) -- (-1,2);
\draw [fill, opacity=0.2, blue, path fading = east] (1,0) -- (2,1) -- (2,2) -- (1,2) -- (0,1) -- (1,0);
\draw [fill, opacity=0.2, blue, path fading = north] (1,0) -- (2,1) -- (2,2) -- (1,2) -- (0,1) -- (1,0);
\draw [fill, opacity=0.2, blue, path fading = west] (-1,0) -- (-2,1) -- (-2,2) -- (-1,2) -- (0,1) -- (-1,0);
\draw [fill, opacity=0.2, blue, path fading = north] (-1,0) -- (-2,1) -- (-2,2) -- (-1,2) -- (0,1) -- (-1,0);
\draw [fill, opacity=0.2, blue, path fading = east] (0,-1) -- (3,-4) -- (4,-4) -- (4,-3) -- (1,0) -- (0,-1);
\draw [fill, opacity=0.2, blue, path fading = south] (0,-1) -- (3,-4) -- (4,-4) -- (4,-3) -- (1,0) -- (0,-1);
\draw [fill, opacity=0.2, blue, path fading = west] (0,-1) -- (-3,-4) -- (-4,-4) -- (-4,-3) -- (-1,0) -- (0,-1);
\draw [fill, opacity=0.2, blue, path fading = south] (0,-1) -- (-3,-4) -- (-4,-4) -- (-4,-3) -- (-1,0) -- (0,-1);
\draw [-latex, thick, black] (3,-2) -- (3.7,-2); \draw [thick, black] (3.7,-2) -- (4,-2);
\draw [dashed, black] (3,-2) -- (1,-2);
\draw [-latex, thick, black] (-1,-2) -- (0.2,-2); \draw[thick, black] (0.2,-2) -- (1,-2);
\draw [dashed, black] (-1,-2) -- (-3,-2);
\draw [-latex, thick, black] (-4,-2) -- (-3.4,-2); \draw [thick, black] (-3.4,-2) --  (-3,-2);
%
%
%
\node [anchor = north east] at (-.8,-.8) {$a'$};
\node [anchor = north west] at (.8,-.75) {$b'$};
\node 
at (-1,1) {$b''$};
\node 
at (1,1) {$a''$};
\node [anchor = south west] at (-3,-3) {$a$};
\node [anchor = south east] at (3,-3) {$b$};
\node [anchor = south east] at (-3,-2) {$\Phi$};
\node [anchor = north west] at (-1,-2) {$\overline{\Omega}$};
\node [anchor = north east] at (1,-2) {$\Phi$};
\node [anchor = south west] at (3,-2) {$\overline{\Omega}$};
\node [anchor = north east] at (-4,-3) {$x$};
\node [anchor = north east] at (-3,-4) {$x'$};
\node [anchor = north west] at (3,-4) {$y$};
\node [anchor = north west] at (4,-3) {$y'$};
\node [left] at (-4,-2) {$z$};
\node [right] at (4.7,-1) {$=\quad$};
\end{tikzpicture}
\begin{tikzpicture}[scale=0.8]
\draw [-open triangle 45, thin, black]  (1,-2)-- (-4,3);
\draw [-open triangle 45, thin, black] (2,-1) -- (-3,4);
\draw [-open triangle 45, thin, black]  (-2,-1)-- (3,4);
\draw [-open triangle 45, thin, black] (-1,-2) -- (4,3);
\draw [fill, opacity=0.2, blue, path fading = east] (1,0) -- (4,3) -- (4,4) -- (3,4) -- (0,1) -- (1,0);
\draw [fill, opacity=0.2, blue, path fading = north] (1,0) -- (4,3) -- (4,4) -- (3,4) -- (0,1) -- (1,0); 
\draw [fill, opacity=0.2, blue, path fading = west] (-1,0) -- (-4,3) -- (-4,4) -- (-3,4) -- (0,1) -- (-1,0);
\draw [fill, opacity=0.2, blue, path fading = north] (-1,0) -- (-4,3) -- (-4,4) -- (-3,4) -- (0,1) -- (-1,0);
\draw [fill, opacity=0.2, blue, path fading = east] (0,-1) -- (1,-2) -- (2,-2) -- (2,-1) -- (1,0) -- (0,-1);
\draw [fill, opacity=0.2, blue, path fading = south] (0,-1) -- (1,-2) -- (2,-2) -- (2,-1) -- (1,0) -- (0,-1);
\draw [fill, opacity=0.2, blue, path fading = west] (0,-1) -- (-1,-2) -- (-2,-2) -- (-2,-1) -- (-1,0) -- (0,-1);
\draw [fill, opacity=0.2, blue, path fading = south] (0,-1) -- (-1,-2) -- (-2,-2) -- (-2,-1) -- (-1,0) -- (0,-1);
\draw [-latex, thick, black] (3,2) -- (3.7,2); \draw [thick, black] (3.7,2) -- (4,2);
\draw [dashed, black] (3,2) -- (1,2);
\draw [-latex, thick, black] (-1,2) -- (0.2,2); \draw[thick, black] (0.2,2) -- (1,2);
\draw [dashed, black] (-1,2) -- (-3,2);
\draw [-latex, thick, black] (-4,2) -- (-3.4,2); \draw [thick, black] (-3.4,2) --  (-3,2);
%
%
%
\node 
at (-1,-1) {$a$};
\node 
at (1,-1) {$b$};
\node [anchor = south east] at (-.8,.8) {$b'$};
\node [anchor = south west] at (.8,.8) {$a'$};
\node [anchor = north east] at (-2.2,3.2) {$b''$};
\node [anchor = north west] at (2.3,3.2) {$a''$};
\node [anchor = north east] at (-3,2) {$\Phi$};
\node [anchor = south west] at (-1,2) {$\overline{\Omega}$};
\node [anchor = south east] at (1,2) {$\Phi$};
\node [anchor = north west] at (3,2) {$\overline{\Omega}$};
\node [anchor = north east] at (-2,-1) {$x$};
\node [anchor = north east] at (-1,-2) {$x'$};
\node [anchor = north west] at (1,-2) {$y$};
\node [anchor = north west] at (2,-1) {$y'$};
\node [left] at (-4,2) {$z$};
\end{tikzpicture}
\end{center}
\caption{A graphical representation of the Yang-Baxter equation \eqref{LLR2}.}\label{LLS-ybe}
\end{figure}
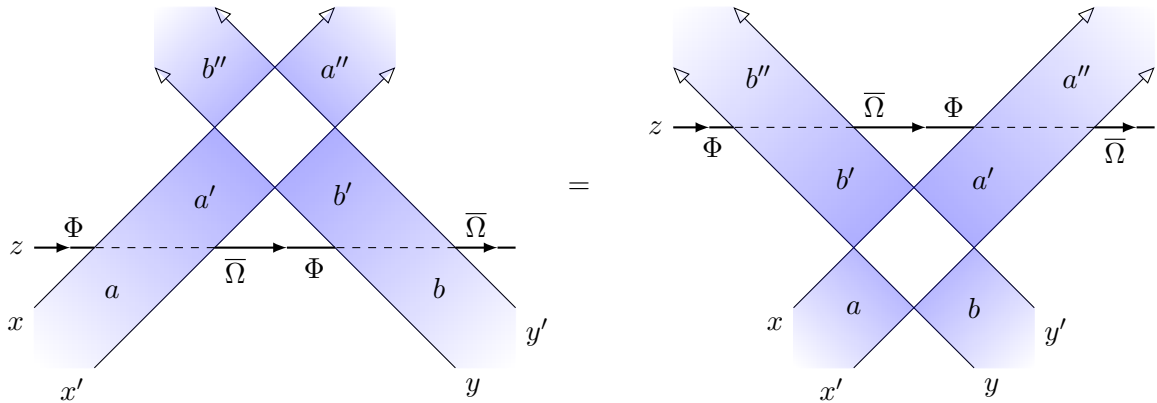

The second exchange relation (arizing upon the substitution of 
\eqref{Sfactor} and \eqref{LL2} into the YBE \eqref{LLR2})
reads 
\beq
\label{psieq2}
\sum_{b} \Phi(x,z)_{a,i}^{b} \ 
\overline{\Omega}(y,z)_{a,i'}^{b}\  
\overline{W}_{x/y}(b-c)
\;=\;
\sum_{b} \overline{W}_{x/y}(a-b)\  \Phi(y,z)_{b,i}^{c} \  
\overline{\Omega}(x,z)_{b,i'}^{c}\;.
\eeq
It can be represented graphically as  
%
\begin{equation}\label{peq2}
\begin{tikzpicture}[scale=0.8,baseline=(current  bounding  box.center)]
\draw [fill, opacity=0.35, blue,path fading=north] (0,2) -- (-1,4) -- (0,4.5) -- (1,4) -- (0,2);
\draw [fill, opacity=0.3, blue,path fading=south] (0,2) -- (1.5,-1) -- (0,-1.5) -- (-1.5,-1) -- (0,2);
\draw [-open triangle 45, black, thin] (-1.5,-1) -- (1,4); \node [above] at (1,4) {$x$};
\draw [-open triangle 45, black, thin] (1.5,-1) -- (-1,4); \node [above] at (-1,4) {$y$};
\draw [dashed, thick] (1,0) -- (0,0); \draw [dashed, thick] (0,0) -- (-1,0);
\draw[-latex, thick] (-2,0) -- (-1.3,0); \draw [thick] (-1.,0) -- (-2,0);  
\draw[-latex, thick] (1,0) -- (1.7,0); \draw [thick] (2.,0) -- (1,0); 
\node [left] at (-2,0) {$i$}; \node [right] at (2,0) {$i'$};
\node [below] at (-1.75,0) {$z$};
\node [below] at (1.6,0) {$z$};
\node at (0,0.66) {$b$}; \node[above] at (0,3.35) {$c$};
\node [below] at (0,-0.3) {$a$};
\node [anchor=south east] at (-1,0) {$\Phi$};
\node [anchor=south west] at (1,0) {$\overline{\Omega}$};
\node [right] at (3.5,1.5) {$=$};
\end{tikzpicture}
\hskip 10mm
\begin{tikzpicture}[scale=0.8,baseline=(current  bounding  box.center)]
\draw [fill, opacity=0.3, blue, path fading=north] (0,1) -- (-1.5,4) -- (0,4.5) -- (1.5,4) -- (0,1);
\draw [fill, opacity=0.35, blue,path fading=south] (0,1) -- (1,-1) -- (0,-1.5) -- (-1,-1) -- (0,1);
\draw [-open triangle 45, black, thin] (-1,-1) -- (1.5,4); \node [above] at (1.5,4) {$x$};
\draw [-open triangle 45, black, thin] (1,-1) -- (-1.5,4); \node [above] at (-1.5,4) {$y$};
\draw [dashed, thick] (1,3) -- (0,3); \draw [dashed, thick] (0,3) -- (-1,3);
\draw[-latex, thick] (-2,3) -- (-1.3,3); \draw [thick] (-1.,3) -- (-2,3);  
\draw[-latex, thick] (1,3) -- (1.7,3); \draw [thick] (2.,3) -- (1,3); 
\node [left] at (-2,3) {$i$}; \node [right] at (2,3) {$i'$};
\node [above] at (-1.75,3) {$z$};
\node [above] at (1.6,3) {$z$};
\node at (0,-0.35) {$a$}; \node at (0,2.34) {$b$};
\node [above] at (0,3.35) {$c$};
\node [anchor=north east] at (-1,3) {$\Phi$};
\node [anchor=north west] at (1,3) {$\overline{\Omega}$};
\end{tikzpicture}
\end{equation} 

\noindent
where the summation over the spins $b$, assigned
to the ``interior'' 
(bounded) faces, is assumed. The boundary spins $a,c$ on the
``exterior'' (unbounded) faces, as well the edge spins $i,i'$ on the
external edges are fixed. 
Finally, with the same graphical notations the
YBE \eqref{LLR2} is represented as in Fig.\ref{LLS-ybe}.
\bigskip\smallskip

It is fairly simple to verify that the relations 
\eqref{psieq2} and \eqref{psieq1}
imply eq.\eqref{LLR2}. First, substitute \eqref{Sfactor} and
\eqref{LL2} into \eqref{LLR2}. Then, using 
(i) the relation \eqref{psieq1}, (ii) the relation \eqref{psieq2}  (twice) 
and, finally, (iii) the relation \eqref{psieq1} again, 
one can easily
transform one side of \eqref{LLR2} into the other. 
To visualize this calculation consider the diagram on the left side 
Fig.~\ref{LLS-ybe}.  Now, let us move 
the horizontal $z$-line upwards, through the intersection points of the
other lines, and then consecutively use an appropriate relation \eqref{psieq1} 
or \eqref{psieq2} at each crossing transition.
In this way the left diagram
in Fig.~\ref{LLS-ybe} is transformed to the right one, thereby proving
\eqref{LLR2}.

\subsection{Calculation of the two-spin weights}
Thus, we have shown that the YBE \eqref{LLR2} with the factorized matrix 
$\bm{{\cal S}}$, given by \eqref{Sfactor}, is a corollary of the
Yang-Baxter type exchange relations \eqref{psieq1} and \eqref{psieq2}.
The next step   
is to solve these relations together with the inversion relations
\eqref{unit1a} and \eqref{unit2a} to find the two-spin weights
$W_x(a-b)$ and $\overline{W}_x(a-b)$. The corresponding 
calculations are presented in the Appendix~\ref{appA}. We show 
that, to within overall normalization
factors and some trivial equivalence transformations\footnote{All the
  equations \eqref{psieq1}, \eqref{psieq2} and \eqref{unit1a},
  \eqref{unit2a} 
(with $\bm{{\cal S}}$ given by \eqref{Sfactor})
are invariant w.r.t a simple transformation   
$\overline{W}_x(n)\to(-1)^n\,\overline{W}_x(n)$ (and, similarly, for
$W$). Our choice of these signs is governed by the positivity
requirements, see the paragraph containing \eqref{physdom2}.}\,, 
the above relations uniquely determine
the two-spin weights
\begin{equation}\label{weights-V}
W_x(n)\;=\;x^{n}\,\frac{(-\qs^{1+n}\,x^2;\qs^2)_\infty}
{(-\qs^{1+n}/x^{2};\qs^2)_\infty}\;, 
\qquad\quad
\overline{W}_x(n)\;=\;W_{\xi/x}(n)\;=\;\left(\frac{\iq}{x}\right)^{n}\,
\frac{(\qs^{2+n}/x^{2};\qs^2)_\infty}{(\qs^n\,x^2;\qs^2)_\infty}\;,
\end{equation}
where $\xi=\ri\sqrt{\qs}$ and $(x;\qs^2)_\infty$ denotes the
$\qs$-Pochhammer symbol \beq (x;\qs^2)_\infty=\prod_{k=0}^\infty
\,(1-x\,\qs^{2k})\,.  \eeq
Below we will use the notations
\beq\label{sigma-def}
\Si(x)=
\frac{(-\qs/x;\qs)_\infty}{(x;\qs)_\infty}\,,\qquad\qquad
\kappa_s=\frac{(\qs;\qs)_\infty}{(-\qs;\qs)_\infty}\,.\qquad
\eeq
The weights \eqref{weights-V}  possess the following important properties:
\begin{enumerate}[(i)]
\item {\sl Positivity}

\noindent
If the parameter $\xi=\ri\sqrt{\qs}$ and the
spectral variable $x$  
are real and belong to the domain  
\beq\label{physdom2}
0<\xi<x<1\,,
\eeq
then the weights \eqref{weights-V} are {\em real and positive}.
\item {\sl Reflection and crossing symmetry}
\beq\label{crossing2}
W_x(n)=W_x(-n)\,,\qquad \overline{W}_x(n)=\overline{W}_x(-n)\,,\qquad
\overline{W}_x(n)=W_{\xi/x}(n)\,,
\eeq
\item {\sl Second order recurrence relations}
\beq\label{recur2}
\frac{\overline{W}_x(n)}{\overline{W}_x(n-2)}\;=\;
-\,\frac{\,\qs-\qs^{n-1}\,x^2}{x^2-\qs^n}
\;,\qquad
\frac{{W}_x(n)}{{W}_x(n-2)}\;=\;
\frac{x^2+\qs^{n-1}}{1+\qs^{n-1} x^2}\,,\qquad n\in{\mathbb Z}\,,
\end{equation}
\item {\sl Inversion relations} 
\begin{equation}\label{inv1}
\sum_{n\in\mathbb{Z}}\; \overline{W}_x(a-n)\, 
\overline{W}_{1/x}(n-b)\;=\;\Upsilon(x)\, \delta_{a,b}\,,
\qquad
W_x(n)\,W_{1/x}(n)\;=\;1\,.
\end{equation}
where 
\beq
\Upsilon(x)=\kappa_s^2\, \Si(x^2)\,\Si(x^{-2})\,,
\eeq
with $\Si(x)$ and $\kappa_s$ defined in \eqref{sigma-def}.
Using the crossing symmetry \eqref{crossing2} 
these relations can also be rewritten as 
\begin{equation}\label{inv2}
\sum_{n\in\mathbb{Z}}\; {W}_{\xi/x}(a-n)\, 
W_{\xi x}(n-b)\;=\; 
\Upsilon(x)\, \delta_{a,b}\,,
\qquad
\overline{W}_{\xi/x}(n)\,\overline{W}_{\xi x}(n)\;=\;1\,.
\end{equation}

\item
{\sl Star-triangle relation}

\noindent
The weights \eqref{weights-V} 
satisfy the star-triangle relation of
the form,
\beq
\begin{array}{l}
\ds \sum_{d\in {\mathbb Z}}\;
\overline{W}_{y/z}(a-d)\;W_{x/z}(b-d)\;\overline{W}_{x/y}(d-c)\,
\\[.5cm]
\phantom{W_{\xi/x}(a-d)\,W_{\xi/y}(b-d)\,}
={\cal R} \;W_{x/y}(b-a)\;\overline{W}_{x/z}(a-c)\;W_{y/z}(b-c)\,,
\end{array}\label{str-main2}
\eeq
where ${\cal R}$ is the scalar factor 
\beq
{\cal R}=\kappa_s\,\frac{\Si(y^2/z^2)\,\Si(x^2/y^2)}{\Si(x^2/z^2)}\,,
\eeq
and the function $\Si(x)$ and the constant $\kappa_s$ are 
defined in \eqref{sigma-def}. Note that $\kappa_s$ is the
``rapidity-idependent factor'' of the star-triangle
relation \cite{Bax02rip}, 
which does not depend on the normalization of the weights
$W$ and $\overline{W}$, obeying the crossing symmetry \eqref{crossing2}.   
\end{enumerate}

\begin{figure}[ht]
\begin{center}
\includegraphics[scale=0.39]{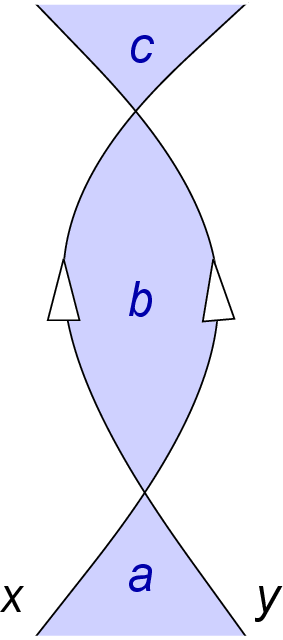}
\hskip 1em
\includegraphics[scale=0.44]{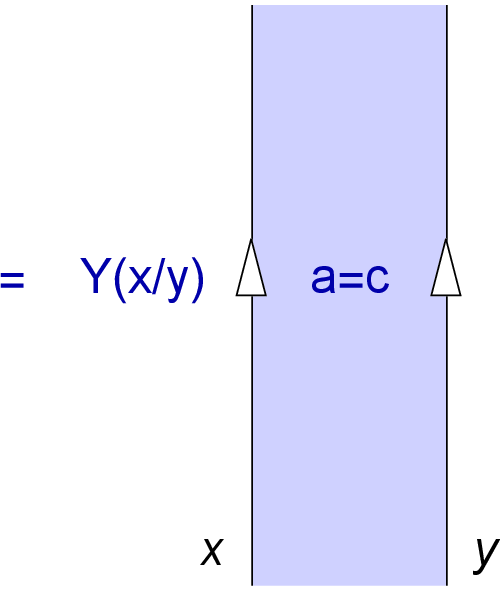}
\hskip 6em
\includegraphics[scale=0.4]{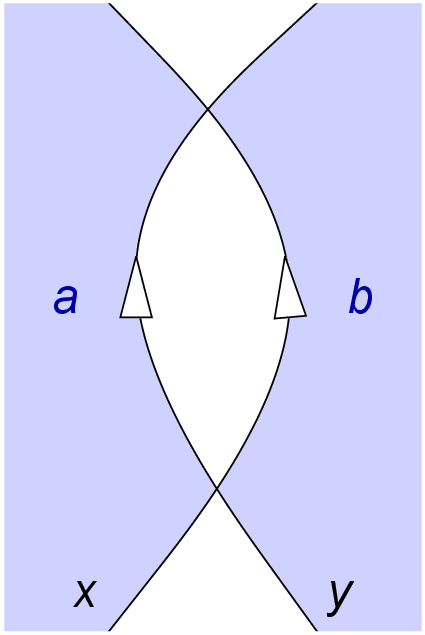}
\hskip 1em
\includegraphics[scale=0.4]{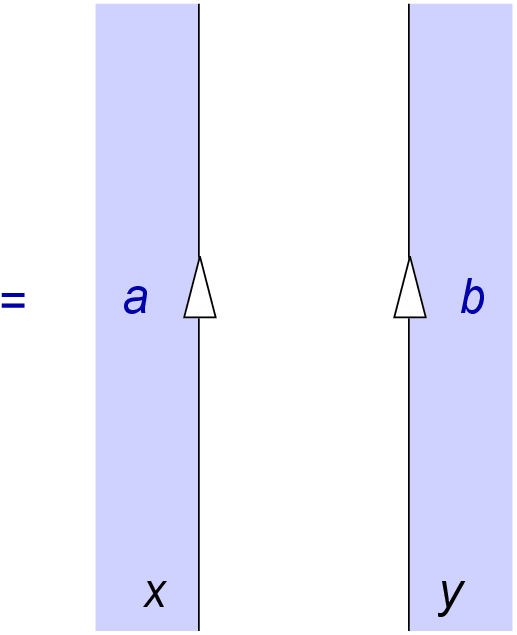}
\end{center}
\caption[  ]{Graphical representation of the first and second  
inversion relations in \eqref{inv1} (on the left and right sides of
the picture, repectively), using the graphical notations
\eqref{W-shade} for the Boltzmann weights.  }
\label{fig-unit12}
\end{figure}

\noindent
Using the graphical notations \eqref{W-shade} one can represent 
the inversion relations \eqref{inv1} as in Fig.\ref{fig-unit12}. 

\begin{figure}[ht]
\begin{center}

\begin{tikzpicture}[scale=.9,baseline=(current  bounding  box.center)]
\draw [-open triangle 45, thin] (2,-1) -- (2,3); 
\draw [-open triangle 45, thin] (-1,.5) -- (3,2.5);
\draw [-open triangle 45, thin] (3,-.5) -- (-1,1.5);
\node [below] at (2,-1) {$y$}; 
\node [below] at (-1,0.5) {$x$};
\node [right] at (3,-.5) {$z$};
\node [right] at (-1.3,1.05) {$b$};
\node [right] at (2.25,-0.8) {$a$};
\node [right] at (1.,1) {$d$};
\node [right] at (2.25,2.8) {$c$};
\draw [fill, opacity=0.25, blue,path fading=west] 
(0,1) -- (-1,.5) -- (-2,1)--(-1,1.5)--(0,1);
\draw [fill, opacity=0.25, blue] 
(2,0) -- (2,2) -- (0,1)--(2,0);
\draw [fill, opacity=0.25, blue,path fading=south] (2,-1) -- (2,0) -- (3,-.5) -- (3,-1.54)--(2,-1);
\draw [fill, opacity=0.25, blue,path fading=north] (2,2) -- (2,3) -- 
(3,3.5) -- (3,2.5) -- (2,2);

\node [right] at (4,1) {$={\ \cal R}$};
\end{tikzpicture}
\hskip 2mm
\begin{tikzpicture}[scale=.97,baseline=(current  bounding  box.center)]
\draw [-open triangle 45, thin] (0,-1) -- (0,3.); 
\draw [-open triangle 45, thin] (-1,-0.5)--(3,1.5); 
\draw [-open triangle 45, thin] (3,0.5)--(-1,2.5); 
\draw [fill, opacity=0.25, blue,path fading=east]
(0,3) --(1.5,2.5)--(3,1.5) -- (2,1) -- (0,2);
\draw [fill, opacity=0.25, blue,path fading=west]
(0,0) -- (-1,-.5)--(-1.5,1)--(-1,2.5) -- (0,2.);
\draw [fill, opacity=0.35, blue,path fading=east]
(0,-1) -- (0,0) -- (2,1) -- (3,.5) -- (1.5,-.5)-- (0,-1);

\node[above] at (0,3) {$\phantom{y}$};
\node [below] at (0,-1) {$y$};


\node [below] at (3,.5) {$z$};
\node [below] at (-0.7,-0.4) {$x$};
\node [above] at (1.3,1.8) {$c$};
\node [below] at (1.4,0.3) {$a$};
\node [right] at (-1,1) {$b$};
\end{tikzpicture}
\end{center}
\caption
{Graphical representation of the star-triangle relation \eqref{str-main2}.}
\label{fig_str}
\end{figure}
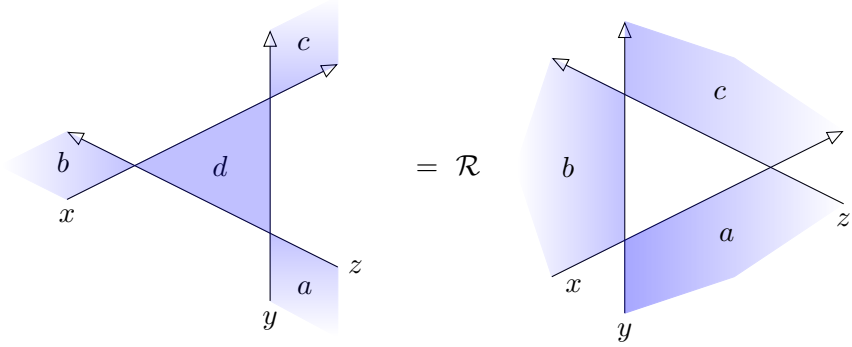

\noindent
Similarly, the
star-triangle relation \eqref{str-main2} is presented in Fig.\ref{fig_str}. 
Recall, that in these figures the spins are assigned to the shaded faces,
the summation over the interior spins is assumed and 
the boundary spins are kept fixed. 
The proof of these relations is given 
below. Here we just remark  
that the properties (iii), (iv), listed above, 
ensure the fullfilment of all defining relations for the 
weights \eqref{weights-V}. Indeed, 
using the explicit form of the face-vertex vectors \eqref{Phi-def} and
\eqref{Omega-def} one can simply rewrite the exchange relations 
\eqref{psieq1}, \eqref{psieq2} as
the recurrence relations \eqref{recur2}. Next, the 
inversion relations \eqref{unit1a} and \eqref{unit2a} simply follow from the
\eqref{Sfactor}, \eqref{inv1} and \eqref{inv2}. 

\subsection{Duality transformation}
Since the two-spin weights only depend on the spin difference, 
is natural to consider their Fourier transformation 
\begin{equation}\label{Fourier0}
\overline{\Om}_{x}(\varphi)\;=\;{\kappa_s}^{-1}{\Si(x^2)}\, 
\sum_{n\in{\mathbb Z}}\re^{\ri\varphi n} \, {W}_{x}(n)
\,,
\qquad
\Om_x(\varphi)\;=\;\, \kappa_s^{-1}\,\Si(x^2)^{-1}\,
\sum_{n\in{\mathbb Z}}\, \re^{\ri\varphi n}\, \overline{W}_x(n)\,, 
\end{equation}
with real $0\le\phi<2\pi$. From \eqref{crossing2} it follows that 
\begin{equation}\label{sym2}
\Om_x(\varphi)\;=\;\Om_x(-\varphi)\;,\qquad
\overline{\Om}_x(\varphi)\;=\;\overline{\Om}_x(-\varphi)\;,
\qquad \overline{\Om}_{x}(\varphi) \;=\; \Om_{\iq/x} (\varphi)\,.
\end{equation}
Explicitly, one obtains
\begin{equation}\label{Fourier1}
\begin{array}{rcll}
\overline{\Om}_{x}(\varphi)\;&=&\;
\Si(x\re^{\ri \phi})\,\Si(x\re^{-\ri \phi})
\;&=\;\ds\frac{\ds (-q\,x^{-1}\,\EXP^{\ii\varphi},-q\,x^{-1}\,\EXP^{-\ii\varphi};q)_\infty}
{\ds (x\,\EXP^{\ii\varphi},x\,\EXP^{-\ii\varphi};q)_\infty}\;,
\\[.6cm]
{\Om}_{x}(\varphi)\;&=&\;
\Si(\xi x^{-1}\re^{\ri \phi})\,\Si(\xi x^{-1}\re^{-\ri \phi})
\;&=\;\ds
\frac{\ds (\iq\,x\,\EXP^{\ii\varphi},\iq\,x\,\EXP^{-\ii\varphi};q)_\infty}
{\ds (\iq\,x^{-1}\,\EXP^{\ii\varphi},\iq\,x^{-1}\,\EXP^{-\ii\varphi};q)_\infty}\,.
\end{array}
\end{equation}
The derivation is
based on an extension of the 
Ramanujan ${}_1\psi_1$ bilateral summation formula \cite{Vildanov:2012},
\begin{equation}\label{rama}
\sum_{n\in\mathbb{Z}} \frac{(b\,q^n;q^2)_\infty}{(a\,q^n;q^2)_\infty}\, z^n 
\;=\;
\frac{(q;q)_\infty}{(-q;q)_\infty}
\,
\frac{(b/a;q^2)_\infty}{(b/az^2;q^2)_\infty}
\,
\frac{(-q/z;q)_\infty}{(z;q)_\infty}
\,
\frac{(az,q/az;q)_\infty}{(a,q/a;q)_\infty}
\end{equation}
valid for $\sqrt{|b/a|}<z<1$. 

Using \eqref{Fourier0} and \eqref{Fourier1} it is easy to prove the first relation in \eqref{inv1} 
(the second one there is obvious) and also derive the inversion relations for 
the $\Om$-weights,
\beq\label{inv4_new}
\frac{1}{(2\pi)^2}\int_{-\pi}^{\pi} d\varphi \, \overline{\Om}_x(\alpha-\varphi) \,\overline{\Om}_{1/x}(\varphi-\beta) \,=\, \kappa_s^{-2}\, 
\Si(x^2)\,\Si(x^{-2}) \, \delta(\alpha-\beta)\,,\qquad
\Om_x(\varphi) \, \Om_{1/x}(\varphi)\,=\,1\,.
\end{equation}
Finally, equivalently rewriting 
the star-triangle relation \eqref{str-main2} in terms of 
the $\Om$-weights \eqref{Fourier0}, one obtains
\beq
\begin{array}{l}
\ds\int_{-\pi}^\pi\frac{\rd \varphi}{2\pi}\,\,
\overline{\Om}_{y/z}(\alpha-\varphi)\;\Om_{x/z}(\beta-\varphi)\;
\overline{\Om}_{x/y}(\varphi-\gamma)\,
\\[.5cm]
\ds\phantom{\Om_{\xi/x}(a-d)\,\Om_{\xi/y}(b-d)\,}
=\ds\widetilde{{\cal R}} \;\Om_{x/y}(\beta-\alpha)\;
\overline{\Om}_{x/z}(\alpha-\gamma)\;\Om_{y/z}(\beta-\gamma)\,,
\end{array}\label{str-fourier}
\eeq
where 
\beq\label{Rtilda}
\widetilde{{\cal R}}=\kappa_s^{-1}\,
\frac{\Si(y^2/z^2)\,\Si(x^2/y^2)}{\Si(x^2/z^2)}\,.
\eeq
Using now the explicit expressions \eqref{Fourier1} one can easily 
transform \eqref{str-fourier} 
into the concise integral identity \eqref{beta-int}, 
presented in the Introduction. As remarked before this identity 
can be derived from the {\em ``constant beta
pentagon equation''
},\  obtained by Garoufalidis and Kashaev 
\cite{Garoufalidis:2017xah}  (see eq.(47) therein). Mathematically, the 
derivation is not very difficult, so we leave it as
an exercise for the reader. However, from a conceptual point of view 
it is very interesting that the {\em pentagon equation} of
\cite{Garoufalidis:2017xah} (which has a specific 
structure of the ``five term quantum dilogarithm identity''
\cite{Faddeev:1993rs} or the ``restricted star-triangle relation'' of
\cite{Bazhanov:1992jq}) can be reinterpreted as the full star-triangle relation
\eqref{str-fourier}. 

To summarize, the proof of the inversion relations \eqref{inv1},
\eqref{inv4_new} is based on the summation formula \eqref{rama}; the
star-triangle relations \eqref{str-main2}, \eqref{str-fourier} and
\eqref{beta-int} can
be derived from the result of \cite{Garoufalidis:2017xah}.
In addition, we have thoroughly verified all key equations numerically.

\section{An integrable model on general planar graphs}
\subsection{``Z-invariant'' lattice models\label{ordef}}
A solvable edge-interaction model on rather general 
planar graphs can be defined in the following way \cite{Bax1,Bax2}.
\begin{figure}[ht]
\begin{center}
\includegraphics[scale=0.75]{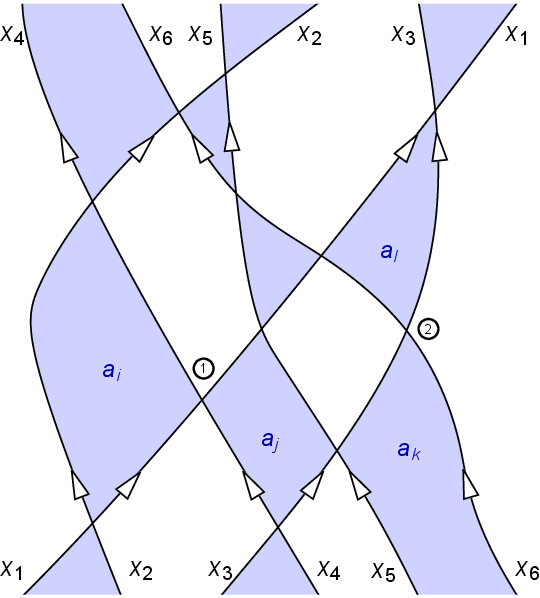}
\end{center}
\caption[The graph $\mathscr{L}$ formed by thin directed lines 
and alternatively shaded faces.]
{The graph $\mathscr{L}$ formed by directed thin lines 
and alternatively shaded faces.
The lines of ${\mathscr L}$
are assigned with the spectral variables $x_1,x_2,x_3,\ldots$.
The symbols $\circled{$\scriptstyle{1}$}$
and $\circled{$\scriptstyle{2}$}$
mark generic examples of the first and second type vertices,
respectively, as defined in \eqref{W-shade}.}
\label{fig-net2}
\end{figure}
Consider a finite set of $L$ directed lines 
forming a graph $\mathscr{L}$ of the type shown in Fig.~\ref{fig-net2}.
The lines (in this case six) head generally from
the bottom of the graph to the top, intersecting one another on
the way. 
Let $V({\mathscr L})$ denote the set of vertices
of ${\mathscr L}$ formed by these intersections.
The lines can go locally downwards, but there can be no
closed directed paths in $\mathscr{L}$. This means that one can
always distort $\mathscr{L}$, without changing its topology, so
that the lines always head upwards.

\begin{figure}[ht]
\begin{center}
\includegraphics[scale=0.75]{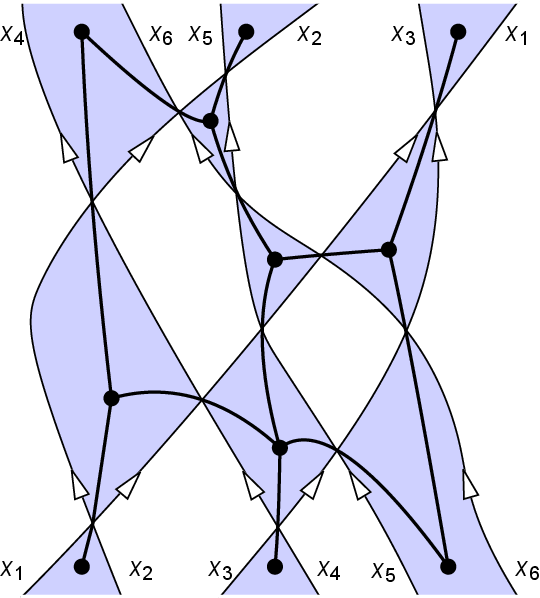}
\end{center}
\caption{The graph ${\mathscr G}$, whose
sites (shown by filled circles) are identified with the shaded faces 
of ${\mathscr L}$, and 
the edges (shown by bold lines) 
are identified with the vertices of ${\mathscr L}$.}
\label{fig-graph-G}
\end{figure}

To each line $\ell$ of ${\mathscr L}$
associate its own spectral variable $x_{\ell}$, taking positive real 
values.
Next, shade alternative faces of $\mathscr{L}$ as shown in
Fig.~\ref{fig-net2} and place integer valued spins $a_i\in{\mathbb Z}$ on all
shaded faces. Among those we will sometimes distinguish 
the interior (bounded) and exterior (unbounded) faces.
The spins $a_i$ and $a_j$ interact only if the corresponding 
faces $i$ and $j$ have a common vertex $(i,j)$.  
There are two types of vertices distiguished  
by the orientation of the lines passing though the vertex 
relative to the shaded faces. 
They are shown in \eqref{W-shade} on the left
(first type) and on the right (second type) sides of the picture. 
It is convenient to introduce a ``spectral parameter ratio variable''
\beq\label{difvar}
s=\left\{\begin{array}{ll}
x/y,\qquad \ \mbox{for a first type vertex}\,,\\[.2cm]
(\xi y)/x,\quad \mbox{for a second type vertex}\,,
\end{array}\right.
\eeq
where the variables $x$ and $y$ are arranged exactly as in \eqref{W-shade}.
Then, each vertex $(i,j)$ is assigned with the Boltzmann weight
$\rho(s_{ij})\,W_{s_{ij}}(a_i-a_j)$, where $a_i$ and $a_j$ 
are the spins on the shaded faces across
the vertex, $s_{ij}$ is the corresponding ratio variable
\eqref{difvar} and $\rho(s_{ij})$ is the normalization factor.
The partition function is defined as a sum over all configurations of  
interior spins with the weight equal to the product of the local
weights over all vertices of ${\mathscr L}$,
\beq\label{z-general}
Z=\sum_{\rm{interior \atop spins}}\quad \prod_{(i,j)\in V({\mathscr L})}
\rho(s_{ij})\,W_{s_{ij}}(a_i-a_j)\,.
\eeq
The spins on exterior faces are kept fixed. 
It is worth noting,
that, more conventionally, the above model can be viewed as an edge
interaction model on the irregular planar graph ${\mathscr G}$, shown
in Fig.~\ref{fig-graph-G} with bold lines and filled circles. The 
sites of ${\mathscr G}$ 
are identified with the (shaded) faces of the original graph 
${\mathscr L}$, while
its edges are identified with the vertices of ${\mathscr L}$, so that 
the set of edges of the new graph ${\mathscr G}$ 
coincides with set of vertices of
the original graph ${\mathscr L}$, i.e., 
$E({\mathscr G})\equiv V({\mathscr L})$. 
With this correspondence the original graph ${\mathscr L}$ is the 
{\em  medial} graph of ${\mathscr G}$. Actually, we could have
started with the planar graph ${\mathscr G}$ and then constructed
${\mathscr L}$ as the medial graph. For instance, taking into account the  
crossing symmetry \eqref{crossing2}, one could easily see that the
expression \eqref{z-main} is just a particular case of
\eqref{z-general} when the graph ${\mathscr G}$ is a homogeneous
square lattice.

The partition function \eqref{z-general} possesses remarkable invariance
properties.
It remains unchanged (up to simple $\rho(s_{ij})$, $\Si(s^{\pm2}_{ij}) $ 
and $\kappa_s$ factors)
by continuously deforming the lines of ${\mathscr L}$ with their boundary
positions kept fixed, as long as the graph ${\mathscr L}$
remains directed. In
particular, no closed directed paths are allowed to appear%
\footnote{Actually, these restrictions can be removed if one properly defines
``reflected'' spectral variables for downward going lines, see 
Sect.3 of \cite{Bax2} for further details.}
.
It is easy to see that all such transformations
reduce to a  combination of the moves shown in
Fig.~\ref{fig_str} and Fig.~\ref{fig-unit12},
corresponding to the star-triangle \eqref{str-main2} and inversion
relations \eqref{inv1}.
%
Given that the graphs ${\mathscr L}$ and
${\mathscr G}$ can undergo rather drastic
changes, the above invariance statement (called the ``Z-invariance''
\cite{Bax1}) is rather non-trivial. 
In particular, it leads to important factorization properties of the
partition function in the large lattice limit. 
Consider a generic planar graph $\mathscr{G}$ with
a large number of sites, $M$, a large number of edges of the order
of $\sim 2M$ and the number of 
boundary sites of the order of $O(M^{1/2})$.
Assume that the boundary spins are kept fixed.
Then, following \cite{Bax1}, one can show
that the  leading asymptotics of the partition function
\eqref{z-general} at large $M$ has the form \cite{Bax1,Bazhanov:2016ajm}
\begin{equation}
\log{ Z}= M\,\log{\kappa_s}+\sum_{(ij)\in E(\mathscr{G})}
\log \kappa_{e}(s_{ij})+O(\sqrt{M})\,,\label{factor}
\end{equation}
where $\zedge(x)$ is the single-edge contribution and  
$\zsite$ is the single-site contribution to the partition
function in the thermodynamic limit. 
Note that the factor $\zsite$, defined in
\eqref{sigma-def}, is  
the ``rapidity-idependent factor'' \cite{Bax02rip} 
for the star-triangle relation \eqref{str-main2}. 
Remarkably, the factors $\zsite$ and $\zedge(x)$ are universal;
they are independent of the graph ${\mathscr G}$.  

Note, that the original formulation of the $Z$-invariant
models \cite{Bax1} involved lattices ${\mathscr L}$ formed by
arbitrary intersections of straight lines.  Here we follow a
generalized formulation \cite{Bax2} where the straight lines are
replaced by arbitrary curved lines, as it was described above.

\subsection{Inversion relations}
In the large-lattice limit the partition function \eqref{z-general}  
can be calculated using the inversion relation method
\cite{Str79,Zam79, Bax82inv}.  
For example, for a regular square lattice of $M$ sites
there are only two different spectral parameters $x$ and $y$. Correspondingly,
a half of edges will have the ratio variable \eqref{difvar} equal
to $x/y$, while the other half will have it equal to $(\xi y)/x$. Let
\beq\label{z-square}
\kappa^{\rm{(sq)}}(x/y)={{Z}\,}^{{1}/{M}},\qquad M\to\infty,
\eeq
be the partition function
per site in the large lattice limit, where the superscript ``(sq)'' stands
for ``square lattice''. Then, using the symmetry, inversion and
star-triangle relations \eqref{crossing2}, \eqref{inv1}, \eqref{str-main2}  
one can show that \cite{Str79, Zam79, Bax82inv},  
\bea
\kappa^{\rm{(sq)}}(x)\,\kappa^{\rm{(sq)}}(x^{-1})&=&\kappa_s^2\,
\rho(x)\,\rho(x^{-1})\,\rho(\xi x)\,\rho(\xi
x^{-1})\,\Si(x^2)\,\Si(x^{-2})\,,\\[.3cm]
\kappa^{\rm{(sq)}}(x)&=&\kappa^{\rm{(sq)}}(\xi/x)\ . \label{func1}
\eea
Together with an appropriate analyticity assumptions 
the above inversion and symmetry relations uniquely determine $\kappa^{\rm{(sq)}}(x)$ (see
\eqref{pf-sq} below). 

For the Ising-type models these relations could
be further refined. First, 
comparing \eqref{z-square} with \eqref{factor} one concludes
\beq
\kappa^{\rm{(sq)}}(x)=\zsite\,\zedge(x)\,\zedge(\xi/x)\,.\label{kappa-sq}
\eeq
Indeed, there are exactly two edges (one of each type) for each site of a
regular square lattice. Correspondingly, the partition function per
site \eqref{kappa-sq} is a product
of the spectral parameter independent single-site
factor $\zsite$ and two single-edge factors $\zedge(x)$ and
$\zedge(\xi/x)$. 

Next, consider the star-triangle relation
\eqref{str-main2}. It is easy see that the factor ${\mathcal R}$ there can be
absorbed into a rescaling of the weights $\overline{W}$ and
to a redefinition of the sum over the interior spin in the LHS of
\eqref{str-main2},  
\beq
 \overline{W}_x(n)\to \frac{1}{\Si(x^2)}\,\overline{W}_x(n)\,\qquad
\ \sum_{d\in{\mathbb Z}}\ \ \to
\ \ \frac{1}{\kappa_s}\,\sum_{d\in{\mathbb Z}}
\ \ \,
\eeq
Now consider the effect of a star-traingle move (from the triangle to
star) in the expression
\eqref{factor}. Such a move exchanges the edges of the first type with 
edges of the second type (and vice versa). 
For instance, the fist type edge with the weight $W_{x/y}(n)$ is replaced
by the second type edge with the (rescaled) weight
$\overline{W}_{x/y}(n)/\Si(x^2/y^2)$. Taking into account the crossing
symmetry \eqref{crossing2} between the weights $W$ and $\overline{W}$,
the normalization factors $\rho(s_{ij})$ in the definition \eqref{z-general} 
and the fact that the partition function \eqref{factor}  
does not change under the star-triangular move, 
one obtains \cite{Bazhanov:2010kz,
Bazhanov:2016ajm}
\beq\label{inv3}
\frac{\kappa_e(\xi/x)}{\rho(\xi/x)}={\Si(x^2)}\,
\frac{\kappa_e(x)}{\rho(x)}\,.
\eeq
Next, the inversion relation moves \eqref{inv1}, shown in
Fig.~\ref{fig-unit12}, trivially lead to   
\beq\label{inv4}
\zedge(\xi x)\,\zedge(\xi/x)=\rho(\xi x)\,\rho(\xi
x^{-1})\,\Si(x^2)\,\Si(x^{-2})
\,,\qquad\zedge(x)\,\zedge(x^{-1})=\rho(x)\,\rho(x^{-1})\,,
\eeq  
where the normalization factors in \eqref{factor} have been taken into account.
Note, that
the relation 
\eqref{inv3} combined with the first relation in \eqref{inv4} imply
the second relation there. 
Finally, it is easy to see that the above relations together
with \eqref{kappa-sq} immediately imply \eqref{func1}.

\subsection{Factorization of the partition function \label{facpar}}
The partition function \eqref{z-general} depends on the exterior spins
and the spectral variables $x_1,x_2,$ $\ldots,$ $x_L$. Of course, it also
depends on the graph  $\mathscr L$, but only on a relative ordering
(permutation) of the rapidity lines at the boundaries and not on their
arrangement inside  the graph.
Naturally, this graph can be identified with an element of the
permutation group.
Then
the partition function $Z$ can be regarded as the permutation group 
representation matrix, acting non-trivially on the spins at the lower and
upper boundaries. In particular, if the associated permutationf
actorizes into a product of two permutations, where the first one
only acts on the first $K$ spectral variables $x_1,x_2,\ldots,x_K$, while
the second one acts on the remaining variables
$x_{K+1},\ldots,x_{L}$,  then using the $Z$-invariance 
the graph ${\mathcal L}$ can be
transformed into two disjoint graphs. Correspondingly, the
partition function 
factorizes (up to simple $\rho(s_{ij})$, $\Si(s^{\pm2}_{ij})$ 
and $\kappa_s$ factors) into the product of two partition functions,
associated with these two graphs.  

\bigskip
\begin{figure}[ht]
\begin{center}
\raisebox{-.5\height}{\includegraphics[scale=0.84]{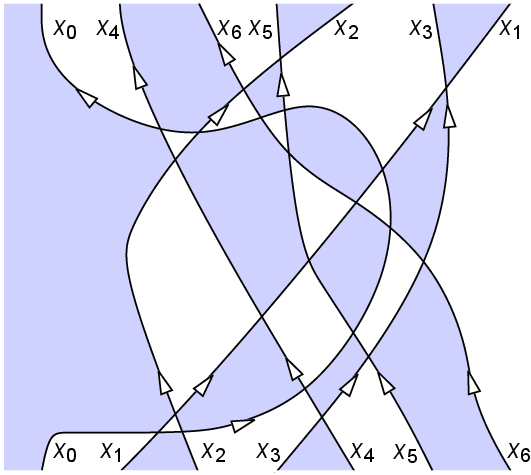}}
$\Longrightarrow\qquad$\raisebox{-.5\height}{\includegraphics[scale=0.84]{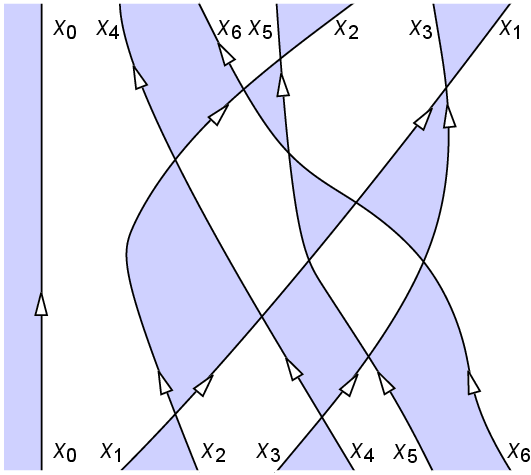}}
\end{center}
\caption[]{Using the star-triangle and inversion relation moves 
the line $x_0$ can be disentangled from the rest of the graph 
${\mathscr L}$. Therefore, the partition functions corresponding to
the above two graphs coincide up to 
the simple $\rho(s_{ij})$, $\Si(s^{\pm2}_{ij})$ 
and $\kappa_s$ factors, associated with the elementary moves. }
\label{fig-net5}
\end{figure}
As an illustration of such factorization consider the graphs shown 
in Fig.~\ref{fig-net5}. Clearly, the permutation, associated with the
graph ${\mathscr L}$ on the left side of the picture, contains the
identity permutation $(x_0)$ (which leaves the first parameter $x_0$ unchanged)
combined with a permutation $(x_4,x_6,x_5,x_2,x_3,x_1)$ of the
remaining six parameters $(x_1,x_2,x_3,x_4,x_5,x_6)$. 
Using the $Z$-invariance, the
leftmost spectral parameter line $x_0$ can be completely disentangled
from the rest of the graph, as shown on the left side of Fig.~\ref{fig-net5}.
Thus, the partition functions for the graphs shown on two
sides of Fig.~\ref{fig-net5} coincide (more precisely,
they differ from each other by
some simple factors, as explained above). Recently, identities of this 
type were intensively studied in the context of the  ``lasso
operator method'' \cite{Chicherin:2017cns} and the ``Yangian
invariance'' of the Zamolodchikov's fishnet diagrams \cite{Zam-fish}
in Quantum Field Theory, see \cite{Kazakov:2023nyu} and references therein. 
Here, we just remark, that such identities could generally be obtained
in a regular way as a consequence of Baxter's $Z$-invariance.
In Appendix~\ref{appB} 
we demonstrate how this works on the example of the $q$-analog of the
${\Dfr}=1$ fishing-net model, obtained in this paper.

\subsection{Dual formulation of the model}
The definition of the dual model follows the same steps as in
Sec.~\ref{ordef}, except that the spin variables are placed on the
unshaded faces of the graph $\mathscr L$. Namely, each unshaded face
is assigned with a continuous spin variable $0\le\varphi_i<2\pi$.
As before, introduce the spectral parameter ratio variable $s_{ij}$ 
by the same formula \eqref{difvar}. 
Then each vertex $(ij)\in V({\mathscr L})$ is assigned
with the Boltzmann weight
$\rho(s_{ij})\,\overline{\Om}_{s_{ij}}(\varphi_i-\varphi_j)$, where 
$\varphi_i$ and $\varphi_j$ 
are the spins on the unshaded faces across
the vertex, the weights $\Om_x(\varphi)$ and
$\overline{\Om}_x(\varphi)=\Om_{\xi/x}(\varphi)$ 
are defined in \eqref{Fourier1}
and $\rho(s_{ij})$ is the normalization factor. In graphical notations
one has 
\beq
(i):
\quad 
\begin{tikzpicture}[scale=.9,baseline={(0,.8)}]
\draw [fill, opacity=.3,blue,path fading=east] 
(1,1) -- (2,0) -- (3,1) --(2,2);
\draw [fill, opacity=.3,blue,path fading=west] 
(0,0) -- (1,1) -- (0,2) --(-1,1);
\draw [-open triangle 45, black, thin] (0,0) -- (2,2);
\draw [-open triangle 45, black, thin] (2,0) -- (0,2);
%
\node [above] at (0,2) {$y$};
\node [above] at (2,2) {$x$};
\node  at (1,2) {$\varphi_j$};
\node  at (1,0) {$\varphi_i$};
\node [right] at (3.1,1) {$
\ds =\;{\overline{\Om}_{x/y}(\varphi_i-\varphi_j)}\;,
$};
\end{tikzpicture}
\quad
(ii):
\quad
\begin{tikzpicture}[scale=.9,baseline=(current  bounding  box.center)] 
\draw [fill, opacity=.3,blue,path fading=north] 
(0,2) -- (1,1) -- (2,2) --(1,3);
\draw [fill, opacity=.3,blue,path fading=south] 
(0,0) -- (1,-1) -- (2,0) --(1,1);
\draw [-open triangle 45, black, thin] (0,0) -- (2,2);
\draw [-open triangle 45, black, thin] (2,0) -- (0,2);
%
\node [above] at (2,2) {$x$};
\node [above] at (0,2) {$y$};
\node at (0,1) {$\varphi_i$};
\node at (2,1) {$\varphi_j$};
\node [right] at (2.1,1) {$
\ds \ \;=\;{\Om}_{x/y}(\varphi_i-\varphi_j)\;,
$};
\end{tikzpicture} 
\eeq
for the first (i) and second (ii) type of vertices, respectively.
The partition function of the dual model 
is defined as an integral over all configurations of  
interior spins with the weight equal to the product of the local
weights over all vertices of ${\mathscr L}$,
\beq\label{z-dual}
Z^{(D)}=\int_0^{2\pi}\cdots\int_0^{2\pi}\prod_{\rm{interior \atop spins}} 
\frac{\rd \varphi_i}{2\pi} \quad \prod_{(i,j)\in V({\mathscr L})}
\rho(s_{ij})\,\overline{\Om}_{s_{ij}}(\varphi_i-\varphi_j)\,.
\eeq
The spins on exterior faces are kept fixed. 

Using the standard arguments \cite{Baxter:1982zz} based
on the duality transformation $\eqref{Fourier0}$ one can
relate \eqref{z-general} with \eqref{z-dual}. In particular, for a
large lattice ${\mathscr L}$ with $M$ shaded faces, one obtains
\beq
\log Z^{(D)}=
 \log Z +  \sum_{(i,j)} \log \frac{\Si(s_{ij}^2)}{\kappa_s} + O(\sqrt{M})\,.
\label{relation}
\eeq
where $\log Z$ is given by (\ref{factor}). For the homogeneous square lattice 
the last relation simplifies to 
\beq
\log Z^{(D)}=
 \log Z -  2 M \log {\kappa_s} + O(\sqrt{M})\,.
\label{relation2}
\eeq
due to the identity $\Si(x^2)\Si(-\qs/x^2)=1$.

\section{Partition function}
\subsection{Inversion relation method} 
In this section we calculate the partition function of the model in
the large lattice limit,
using the inversion relations method \cite{Str79, Zam79, Bax82inv}.  
From now on we will adopt the following normalization 
of the Boltzman weights in \eqref{z-general}
\beq\label{norm1}
\rho(x)=\frac{(-q\,x^{-2};q^2)_\infty}{(-q\,x^2;q^2)_\infty}\,,\qquad
\log \rho(x)=\sum_{m=1}^\infty \frac{(-q)^m\,(x^{2m} -x^{-2m})}
{m(1-q^{2m})}\,.
\eeq
It is useful to note, that
\beq
\rho(\xi x)\rho(\xi/x)=(1-x^2)(1-x^{-2})\,, \qquad 
\rho(x)\rho(x^{-1})=1\,.
\eeq
 With this normalization
the weights $\rho(x) W_x(n)$ and $\rho(\xi/x) \overline{W}_x(n)$,
with $n\in {\mathbb Z}$, are analytic in the ring 
\beq
\xi^2<|x|<\xi^{-1}\,,\qquad 0<\xi<1\,.
\eeq
The factor \eqref{norm1} is chosen such that   
\beq\label{norm2}
\rho(x)\,W_x(0)=\rho(\xi/x)\,\overline{W}_x(0)\equiv1\,.
\eeq

In view of the above, it is natural to assume, 
that for positive real values of $\xi$ 
the edge partition function $\zedge(x)$ is analytic in the ring 
\beq
\xi \re^{-\delta}<|x|<\re^{\delta}\,, \qquad 0<\xi<1\,.
\eeq
where $\delta>0$ is a small, but finite constant. 
Then, the inversion relations \eqref{inv3}, \eqref{inv4} have a unique
solution,
\begin{equation}\label{pf1}
\log \zedge(x)\;=\;\log\rho(x)
-\sum_{m=1}^\infty \frac{(-q)^m\,(x^{2m} -x^{-2m})}
{m(1-q^m)(1+(-q)^m)}\,=\,2\sum_{m=1,3,5,\dots} \frac{q^{2m}(x^{2m}-x^{-2m})}{m(1-q^m)(1-q^{2m})}
\end{equation}
The derivation is straightforward. Taking the logarithms of
\eqref{inv4} and using the Laurent series 
\beq
\log \zedge(x)=\sum_{m=-\infty}^\infty a_m x^{2m}\,,\qquad
\label{Si-exp}
\log\Si(x^2)\;=\; \sum_{m=1}^\infty \frac{x^{2m} - (\xi/x)^{2m}}{m (1-q^m)}
\end{equation}
one obtains a system of a linear equations for the coefficients
$\{a_m\}$, which immediatelly leads to the above result \eqref{pf1}. 
The equation
\eqref{inv3} is automatically satisfied. Mention also a product representation
for \eqref{pf1},
\beq  \label{F-def}
\zedge(x)=\frac{(q\,x^2;q^2)_\infty}{(-q x^2;q^2)_\infty}
\,F(x)\,F(\xi/x)\,,\qquad
F(x)=\prod_{n=1}^\infty\left(\frac{(1+q^{2n} x^2)(1-q^{2n}/x^2)}
{(1-q^{2n} x^2)(1+q^{2n}/x^2)}\right)^n\,.
\eeq

Substituting \eqref{pf1} into \eqref{kappa-sq} one gets the partition
function per site \eqref{z-square} 
for the case of the homogeneous square lattice model
\bea\label{pf-sq}
\log \kappa^{\rm{(sq)}}(x)\;&=&\;\log(\kappa_s\rho(x)\rho(\xi/x)) 
-\sum_{m=1}^\infty 
\frac{
  (x^{2m}+(\xi/x)^{2m})(\xi^{2m}-1)}{m(1-q^m)(1+\xi^{2m})}=\\[.3cm]
%
\;&=&\;\log\kappa_s
+
2\sum_{m=1,3,5,\cdots} \frac{q^m (x^{2m} + (\xi/x)^{2m})}{m (1-q^m)^2}\;,\quad
\qquad \xi^2\;=\;-q\;,
\eea
or in a product form
\beq\label{pf-prod}
\kappa^{\rm{(sq)}}(x)\;=\;\kappa_s 
\,\frac{(q\,x^2,-q^2/x^2;q^2)_\infty}{(-q\,x^2,q^2/x^2;q^2)_\infty}
\,\,F(x)^2\,F(\xi/x)^2\,.
\eeq

The results \eqref{pf1}, \eqref{pf-sq} strongly resemble the expression 
for the partition function of the symmetric 8-vertex model
\cite{Baxter:1972,Baxter:1982zz}. It is, therefore, important to
better understand this connection.

\subsection{Connection to the 8-vertex model}
The $R$-matrix of the symmetric 8-vertex model has the form 
\beq
{\mathcal R}^{(8v)}(u)=\left(\begin{array}{cccc}
\as(u)&&&\dds(u)\\
&\bs(u)&\cs(u)&\\
&\cs(u)&\bs(u)&\\
\dds(u)&&&\as(u)
\end{array}\right)
\eeq
with the Boltzmann weights parameterized as 
\beq\label{abcd-8v}
\begin{array}{rclrcl}
\as(u)&=&\rho_8\,\vartheta_4(\lambda)\,\vartheta_1(\lambda-u)\,\vartheta_4(u)\,,
\qquad\qquad\qquad&  
\bs(u)&=&\rho_8\,\vartheta_4(\lambda)\,\vartheta_4(\lambda-u)\,\vartheta_1(u)\,,\\[.3cm]
\cs(u)&=&\rho_8\,\vartheta_1(\lambda)\,\vartheta_4(\lambda-u)\,\vartheta_4(u)\,,\qquad&
\dds(u)&=&\rho_8\,\vartheta_1(\lambda)\,\vartheta_1(\lambda-u)\,\vartheta_1(u)\,.
\end{array}
\eeq
Here we have used the standard notations for the $\vartheta$-functions
\begin{equation}
\vartheta_1(u)\;=\;\ii \qs^{^1\!\!/\!_4} \EXP^{-\ii u} (\EXP^{2\ii
  u},\qs^2\EXP^{-2\ii u},\qs^2;\qs^2)_\infty\;,\quad 
\vartheta_4(u)\;=\;(\qs\EXP^{2\ii u},\qs\EXP^{-2\ii u},\qs^2;\qs^2)_\infty\;.
\end{equation}
where $\rho_8,\, \lambda,\, u$ and $\qs$ are free parameters of the
model. 
It is convenient to define 
\begin{equation}
\overline{\xi}=\EXP^{\ii\lambda}\,,\qquad x=\EXP^{\ii u}\,,\qquad
\gamma=\ii \qs^{^1\!\!/\!_4}  (\qs^2;\qs^2)_\infty^2 \vartheta_4^{}(0) \;,
\end{equation}
Then in the regime when $x$ and $\overline{\xi}$ are real and positive
and 
\beq
0<\overline{\xi}<x<1\,,\qquad |\qs|<1\,.
\ee
the paritition function per site reads (see Eq.(10.8.44) 
of \cite{Baxter:1982zz})
\begin{equation}\label{kappa-8v}
\log \kappa^{(8)}(x)\;=\;\log(\rho_8\gamma/\overline{\xi})
-\sum_{m=1}^\infty \frac{(x^{2m}+(\overline{\xi}/x)^{2m})(\overline{\xi}^{2m} + (\qs/\overline{\xi}^2)^m)}{m (1-\qs^m) (1+\overline{\xi}^{2m})}\;.
\end{equation}

Let us now identify the parameters $x,\qs, \overline{\xi}$ in \eqref{kappa-8v}
with the corresponding parameters $x,\qs, {\xi}$ in \eqref{pf-sq}.
Thereby we need to set
\beq\label{xi-square}
\re^{2\ri \lambda}=\overline{\xi}^2={\xi}^2=-\qs\,,
\eeq
Then it is not difficult to show that the two partition functions
differ from each other by a simple factor
\begin{equation}
{\kappa^{(8)}(x)}/{\kappa^{\rm(sq)}(x)}\;=
\;\frac{\rho_8\,\gamma\, (-x^2,q x^{-2};q)_\infty}{\xi\,\kappa_s}\,.
\end{equation}
where the normalization factor \eqref{norm1} is taken into account. Thus, the new infinite-state model \eqref{z-main}, introduced in this paper, is 
``weakly-equivalent'' to the 8-vertex model.  
Interestingly, the arising correspondence leads to 
an unphysical regime of the 8-vertex model, where the weight $d(u)$ in \eqref{abcd-8v} is purely imaginary, if $a(u),b(u)$ and $c(u)$ are chosen to be real 
and positive. However, the infinite state-model on the other side of the correspondence has strictly positive Boltzmann weights.

Next, we observe, 
that with the relation \eqref{xi-square} between $\lambda$ and
$\qs$ the
8-vertex model \eqref{abcd-8v} 
reduces to the free fermion model \cite{FanWu,Felderhoff} 
with the condition\footnote{%
In general the weights \eqref{abcd-8v} satisfy the condition 
$$\frac{\as^2+\bs^2-\cs^2-\dds^2}{\as \bs}=-\frac{2\,
\vartheta_4^2\,\vartheta_2(\lambda)\,\vartheta_3(\lambda)}{
\vartheta_2\,\vartheta_3\,\vartheta_4^2(\lambda)}\,.
$$ In the case \eqref{xi-square} the function
$\vartheta_3(\lambda)$ vanishes.
}
\beq
\as^2+\bs^2-\cs^2-\dds^2=0\,,
\eeq
for the Boltzmann weights.
Moreover, it is worth noting that the site factor 
\beq
\kappa_s=\frac{(\qs;\qs)_\infty}{(-\qs;\qs)_\infty^{}}
=(\qs;\qs^2)_\infty^2(\qs^2;\qs^2)_\infty^{}=\vartheta_4(0)=\vartheta_4
\eeq
is just a theta constant. 
The partition function \eqref{kappa-8v} can now be written as a 2D 
free-fermion determinant \cite{FanWu}
\beq\label{kappa-ffm}
\log \kappa^{(8)}=\frac{1}{8 \pi^2}\int_0^{2\pi}
\int_0^{2\pi}\rd\phi_1 \rd\phi_2\,
\log\Big|2 A+2 D\,\cos(\phi_1-\phi_2)+2 
E\,\cos(\phi_1+\phi_2)\Big|
\eeq
where
\beq
A=a^2+b^2\,,\qquad D=c^2-a^2\,,\qquad E=d^2-a^2\,,
\eeq
which is simply related to the partition function of the 2D zero-field Ising 
model \cite{Onsager:1944}. The product representations of the type 
\eqref{pf-prod} for the most general free-fermion model and the
equivalent checkerboard Ising model were obtained
in \cite{Bazhanov:1984ji,Baxter:1986df}.

\subsection{Critical point}
Let us now analyse the behavior of the model 
near the critical point $\qs=-1$, where the model exibits a phase transition. 
Let 
\beq\label{paramet}
\xi=\re^{-\epsilon/2}\,,\qquad \qs=-\re^{-\epsilon}\,,\qquad x^2=\re^{-\theta \epsilon},\qquad \epsilon\to+0\,.
\eeq
where $\theta$ is a new parameter, replacing $x$. Then for the
normalized Boltzmann weights \eqref{norm2} 
of the discrete spin model for $\epsilon\to0$, one obtains,
\beq\label{FZweights}
\frac{W_x(2n)}{W_x(0)}=V_\theta(n)+O(\epsilon)\,,\qquad \frac{\overline{W}_x(2n)}{\overline{W}_x(0)}=V_{1-\theta}(n)+O(\epsilon)
\eeq
where $n\in{\mathbb Z}$ 
\beq
V_\theta(n)= \frac{\Gamma(\frac{1+\theta}{2})}{\Gamma(\frac{1-\theta}{2})}
\frac{\Gamma(n+\frac{1-\theta}{2})}{\Gamma(n+\frac{1+\theta}{2})}\;.
\eeq
such that for large $n$ 
\beq\label{Vasymp}
\log V_\theta(n)= \textstyle
\,{-\,\theta}\,\log |n|\,
+O(1)
+O\big(n^{-1}\,\log |n|\big)
\,,\qquad n\gg1\,.
\eeq  
For odd values of spins
\beq\label{oddval}
\frac{W_x(2n+1)}{W_x(0)}= \epsilon^\theta (1+O(\epsilon))\,,\qquad
\frac{\overline{W}_x(2n+1)}{\overline{W}_x(0)}= \epsilon^{1-\theta}(1+O(\epsilon)) \,.
\eeq
For the dual (compact) formulation of the model one obtains,
\beq
\Om_x(\varphi)={\bf w}_\theta(\varphi)+O(\epsilon)
\,,\qquad \overline{\Om}_x(\varphi)={\bf w}_{1-\theta}(\varphi)+O(\epsilon)\,,
\eeq
where 
\beq\label{wfish}
{\bf w}_\theta(\varphi-\varphi')=|2\sin(\varphi-\varphi')|^{-\theta}
\eeq
Note that the weights \eqref{FZweights} and \eqref{wfish} 
describe the $N\to\infty$ 
limit \cite{AuYang99} 
of the Fateev-Zamolodchikov $Z_N$ model \cite{FZ82}, which is
equivalent to ${\Dfr}=1$ Zamolodchikov's fishing-net model 
\footnote{%
The formulation of \cite{Zam-fish} (with continuous spin variables
on the
real line 
${\mathsf{x},\mathsf{x}'}\in{\mathbb R}$) with  the Boltzmann
weights proportional to
$\sim|\mathsf{x}-\mathsf{x'}|^{-\theta}$ 
is connected to \eqref{wfish} 
by the transformation $\ds \mathsf{x}=\cot\varphi,\;
\mathsf{x}'=\cot\varphi'$.}.

With the parameterization \eqref{paramet} 
the function $F(x)$ defined in \eqref{F-def}  
has the following asymptotics when $\epsilon\to0$,  
\beq\label{F-ass}
F(x)=\re^{-\frac{\pi^2}{8\epsilon}\theta}
\,G(\theta)\,\Big(1-\frac{\pi}{\epsilon}\,
\re^{-{\pi^2}/{\epsilon}}\,\sin\pi\theta+O(\epsilon)\Big)
\eeq
where 
\beq\label{G-def}
G(\theta)=\exp
\left\{\frac{1}{\pi}\int_0^{\pi \theta/2} (x\cot x)\,\rd x\,\right\}=
\re^{\frac{1}{2}\theta}\,\prod_{n=1}^{\infty}\left(
\frac{\Gamma\big(n+\frac{\theta}{2}\big)}{
\Gamma\big(n-\frac{\theta}{2}\big)}\re^{-\theta \psi(n)}\right)\,,\qquad G(1)=\sqrt{2}\,,
\eeq
and $\psi(x)=\rd \log\Gamma(x)/\rd x$ is the logarithmic derivative of
the gamma-function.
In writing \eqref{F-ass} we have indicated the most singular
non-analytic correction term, 
though numerically it could be much smaller than
the regular $O(\epsilon)$ term.    
It follows then that 
partition function (\ref{pf-sq}) has the following expansion when
$\epsilon\to0$,  
\beq\label{expan0}
\log \kappa^{(\textrm{sq})}(x) \;=\; \log \kappa^{(\textrm{sq})}_0(\theta) + \log \kappa^{(\textrm{sq})}_{\textrm{reg}}(\theta) + 
\log \kappa^{(\textrm{sq})}_{\textrm{sing}}(\theta)+\dots\;.
\eeq
The leading term is given by
\beq
\kappa^{(\textrm{sq})}_0(\theta)\;=\;
\frac{\Gamma(1-\frac{\theta}{2})\Gamma(\frac{1+\theta}{2})}{\pi} \mathtt{Z}(\theta)^2\;,
\eeq
where
\beq\label{three-forms}
\begin{array}{ll}
\ds \mathtt{Z}(\theta)&=\;\mathtt{Z}(1-\theta)  \ds\;=\;
\prod_{n=1}^\infty \frac{\Gamma(n+\frac{\theta}{2})\Gamma(n+\frac{1-\theta}{2})\Gamma(n-\frac{1}{2})}
{\Gamma(n-\frac{\theta}{2})\Gamma(n-\frac{1-\theta}{2})\Gamma(n+\frac{1}{2})}
\\[.8cm]
& \ds=\;
\sqrt{\cos\frac{\pi\theta}{2}}\,
\exp\left(\frac{1}{8}\int_{\mathbb{R}} \frac{dw}{w} \;\frac{\sinh(2\theta w)}{\cosh^2(w)}\right)\\[.8cm]
&\ds=\;\exp\Big\{\frac{1}{16 \pi^2}
\iint_0^{2\pi}\rd\phi_1 \rd\phi_2\,
\log\Big(2+2\sin^2(\pi\theta/2)\,\cos(\phi_1-\phi_2)-2 
\cos^2(\pi\theta/2)\,\cos(\phi_1+\phi_2)\Big)\Big\}
\end{array}
\eeq
is the partition function per edge for ${\Dfr}=1$ Zamolodchikov's
fishing net model presented in three different forms. Next, the term 
$\log \kappa^{(\textrm{sq})}_{\textrm{reg}}(\theta)$ 
stands for a well defined regular series in positive integer powers of
$\epsilon$ vanishing in the limit $\epsilon\to 0$, and  $\log
\kappa^{(\textrm{sq})}_{\textrm{sing}}$ denotes 
the most 
singular non-analytic contribution near $\epsilon\sim0$ 
\beq\label{mostsing}
\log \kappa^{(\textrm{sq})}_{\textrm{sing}}\;=\; -\frac{4\pi}{\epsilon}\,
\EXP^{-\pi^2/\epsilon}\,\sin\pi \theta\,.
\eeq
The dots in \eqref{expan0} denote less singular terms of the order $O(\widetilde{\qs}^n)$ and $O(\widetilde{\qs}^{n+1}\,\log \widetilde{\qs})$
with $n\ge1$, where $\widetilde{\qs}=\EXP^{-\pi^2/\epsilon}$.

The first line in \eqref{three-forms} is the original result
of \cite{Zam-fish}.  
The second line is (the square root of) the
partition function per site of the six-vertex model at the free-fermion point
and the third line is its free-fermion determinant representation, see
Appendix~\ref{appC} for additional details. As
a side remark note, that the partition function of the general fishing-net
model \cite{Zam-fish}, where the spins are taking values in ${\mathbb R}^\Dfr$ with
$\Dfr\ge1$, 
can be represented as a product of the same determinants
\eqref{three-forms}
with shifted values of the spectral parameter $\theta$. 

\section{Conclusion}
In this paper we obtained a new solution of the star-triangle relation
with positive Boltzmann weights.  The solution is presented to two
equivalent forms: the non-compact form, with spins taking arbitrary
integer values and the compact form, with continuous spins taking
values on the circle $0\le \varphi <2\pi$.  It is commonly accepted
that the solutions of the Yang-Baxter equation (with the star-triange
relation being a particular case) are completely described by the
theory of quantum groups
\cite{Drinfeld:1987,Jimbo:1986a,Faddeev:1987ih}, so finding a new solution  
should, in principle, be reducible to a routine task of the
representation theory.  From the algebraic point of view the new
solution is closely related to the six-vertex model. It is, indeed,
connected with the problem of the construction of an intertwiner for
two particular infinite-dimensional representations of the quantum
affine algebra $U_{\qs}(\widehat{sl}(2))$, which is the simplest and
most well studied quantum affine algebra. However, the fact that this
solution has not so far been discovered indicates that the problems of
the representation theory, perhapse, are not so routine. Indeed, a
more rigorous consideration of the reccurence relations for the
Boltzmann weights \eqref{r2} has lead to non-hypergeometric type
summation formulae, e.g., the star-triangle equation \eqref{str-main}
itself and the generalized Ramanujan ${}_1\psi_1$ summation
formula \eqref{rama}. It would be interesting to explore apllication
of these ideas to other algebras.

We also presented an exactly solvable two-dimesional lattice model
describing interaction of integer-valued spins $a,b,\ldots$ (often
called ``heights'') on the neigbouring sites of the square lattice. It
can be viewed as a discrete solid-on-solid (SOS) model for a surface
roughening transition.
\begin{figure}[ht]
\begin{center}
\includegraphics[scale=1.2]{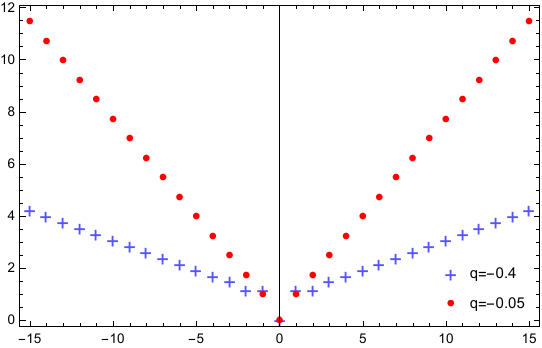}
\caption[Interaction energy]{The two-spin interaction energy 
$\beta E(a,b)
=-\log\big({W_{\sqrt{\xi}}(a-b)}/{W_{\sqrt{\xi}}(0)}\big)$
for the discrete SOS model, as a function of the spin difference $(a-b)$.
The linear growth at large arguments is described by \eqref{linear}.
\label{logwplot}}
\end{center}
\end{figure}
The model have one temperature-like parameter $\qs=-\xi^2$,  
$0<\xi<1$. In the symmetric case (with the same interaction on 
the horisontal and vertical
edges) the two-spin interaction energy multiplied by the inverse
temperature, illustrated in Fig.~\ref{logwplot},
can be approximated by a simplified formula  
\beq\label{linear}
\beta E(a,b)=-\log \frac{W_{\sqrt{\xi}}(a-b)}{W_{\sqrt{\xi}}(0)}\sim 
-\half |a-b|\log \xi\,,\qquad |a-b|\gg1\,.
\eeq
So the parameter $-\log \xi$ plays the role of the inverse
temperature.  At small $\xi$ one expect an ordered state, while at
$\xi\sim 1$ the linear interaction disappears, so the spins are
expected to be disordered (to avoid confusions, note that at $\xi=1$
the formula \eqref{linear} is not applicable, see
Eqs.~\eqref{FZweights}, \eqref{Vasymp} and \eqref{oddval} in the main
text). From the exact result \eqref{pf-sq} for the partition function
of the model it follows, that the most singular contribution to the
free energy near the critical point $\xi=1$, is given
by \eqref{mostsing},
\beq
\log \kappa^{\rm (sq)}_{\rm sing}\sim -\,\frac{\,4\pi}{\epsilon\,}\,\, 
\re^{-\pi^2/\epsilon}\,,
\qquad \xi=\re^{-\epsilon/2},\qquad \epsilon\to0\,,
\eeq
which, with an account of \eqref{FZweights},
indicates an infinite order phase transition.
At the critical point the model reduces 
to the ${\Dfr}=1$ Zamolodchikov's fishing-net model \cite{Zam-fish}.
\begin{figure}[ht]
\begin{center}
\includegraphics[scale=1.2]{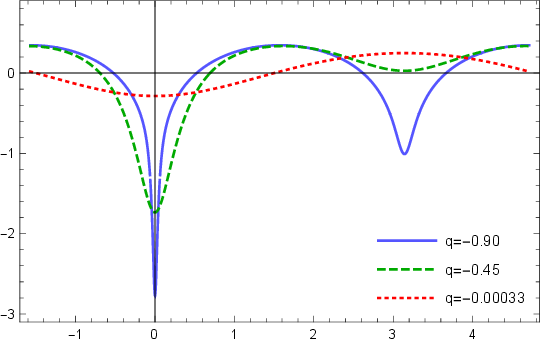}
\caption[Interaction energy]{The two-spin interaction energy 
$\beta E^{(D)}(\varphi)
=-\log \Om_{\sqrt{\xi}}(\varphi)$ as a function of the spin difference 
$\varphi$ in the dual formulation of the model.\label{omegaplot}}
\end{center}
\end{figure}

In the dual formulation the spins take continuous values on the circle
$0\le\varphi_i<2\pi$ and the neighbouring
spin interaction energy $\beta E^{(D)}(\varphi)$
becomes a $2\pi$-periodic and even function of the spin difference 
$\varphi=\varphi_i-\varphi_j$.  It is illustrated in
Fig.~\ref{omegaplot}. For small $\qs=-\xi^2$ the interaction energy is just a 
weakly oscillating function 
\beq
 \beta E^{(D)}(\varphi)=
-\log\Om_{\sqrt{\xi}}(\varphi)=
-2\sqrt{\xi}\cos\varphi+O(\xi)\,,\qquad \xi\sim0\,,
\eeq
but when $\xi\to 1$ it devolopes two sharp and gradually deepening 
minima at $\varphi=0$ and $\varphi=\pi$ (see Fig.\ref{omegaplot} and Eq.\eqref{wfish} for the limiting form of the weight $\Om_x(\varphi)$ at $\xi=1$). 
It would be interesting to verify (for instance, using the approach of 
\cite{Caselle:2019khe}) 
whether the phase transition in the
model could be interpreted as a Berezinskii-Kosterlitz-Thouless
transition \cite{Berezinskii:1971,Kosterlitz:1973} 
induced by contributions of vortex/antivortex configurations
to the partition function. Moreover, it would be useful to calculate
the spin correlation functions, which we postpone to a future
publication.

Finally, mention an intriguing connection of the partition function of 
the model with the partition function of the off-critical 8-vertex free-fermion model. Essentially, the two partition function coincide (more precisely, they differ by a simple factor which can be absorbed into the normalization of the Boltzmann weights). This means that the partition function of the model can be represented as a free-fermion determinant.
It would be extremely interesting to understand this connection 
on the level of the Bethe ansatz. 
 
\section*{Ackknowledgements} The authors thank R.~J.~Baxter,
R.~M.~Kashaev, S.~L.~Lukyanov, J.~H.~H.~Perk, V.~P.~Spiridonov and 
S.~O.~Warnaar 
for very stimulating discussions at various stages of this work.
%
SMS acknowledges the support of the Australian Research Council 
grant DP190103144.

\app{Solution of the recurrence and inversion relations\label{appA}}
Consider a subset of \eqref{recur2}, \eqref{inv1} and \eqref{inv2}
only involving the weight $\overline{W}_x(n)$,
\beq\label{recur2a}
\frac{\overline{W}_x(n)}{\overline{W}_x(n-2)}\;=\;
\left(-\frac{\qs}{x^2}\right)\,\frac{(1-\qs^{n-2}\,x^2)}{(1-\qs^n/x^2)}
\;,\end{equation}
\beq\label{u1a}
\sum_{n\in{\mathbb Z}}\,\overline{W}_x(a-n)\,
\overline{W}_{1/x}(n-b)\simeq \delta_{a,b}\,,\qquad 
\overline{W}_x(n)\,\overline{W}_{-\qs/x}(n)\simeq 1\,,
\qquad a,b,n\in{\mathbb Z}\,,
\eeq
Introduce the notation
\beq
w_x(n)=\left\{\begin{array}{l} w_x(0)\,,\quad n=\mbox{even}\,,\\
w_x(1)\,,\quad n=\mbox{odd}\,. \end{array}\right.
\eeq
where $w_x(0)$ and $w_x(1)$ are two arbitrary functions of $x$.
Then, the most general solution of \eqref{recur2a} for
$\overline{W}_x(n)$ can be written as 
\beq
\overline{W}_x(n)=w_x(n)\,\left(\frac{\xi}{x}\right)^{n}\,\frac{(\qs^{2+n}/x^2;\qs^2)_\infty}
{(\qs^{n}\,x^2;\qs^2)_\infty}\,
\eeq
where $(x;\qs^2)_\infty$ denotes the $\qs$-Pochhammer symbol 
\beq
(x;\qs^2)_\infty=\prod_{k=0}^\infty \,(1-x\,\qs^{2k})\,.
\eeq
Substituting this into the first equation in 
\eqref{u1a} one obtains an infinite set of 
equations of the form
\beq
\begin{array}{l}
A_x(a-b)\, w_x(0)\, w_{1/x}(0) +B_x(a-b)\, w_x(1) \,w_{1/x}(1)=0\,,\qquad
a-b=\pm2,\pm4,\ldots\,,\\[.3cm]
C_x(a-b)\, w_x(0)\, w_{1/x}(1) +D_x(a-b)\, w_x(1)\,
w_{1/x}(0)=0\,,\qquad
a-b=\pm1,\pm3,\ldots
\end{array}
\eeq
where the coefficients $A,B,C,D$ are known function of the spectral variable 
$x$ and the spin difference (they can be explicitly calculated
using the Ramanujan bilateral summation formula ${}_1\psi_1$). 
A carefull analysis shows that all
these equations are consistent and reduce to only two equations 
\beq
\begin{array}{l}
w_x(0)\, w_{1/x}(0)-w_x(1) \,w_{1/x}(1)=0\,,\\[.3cm]
w_x(0)\, w_{1/x}(1)-w_x(1)\, w_{1/x}(0)=0\,,
\end{array}
\eeq
which have a simple solution $w_x(1)=\pm\,w_x(0)$.
Therefore, we will set  
\beq
w_x(1)=w_x(0)=1\,, \qquad \forall n\,,\label{w01}
\eeq
which fixes
$\overline{W}_x(n)$ to within an overall normalization and 
a trivial equivalence transformation $\overline{W}_x(n)\to
(-1)^n\,\overline{W}_x(n)$. Then, using \eqref{recur2a} together with
\eqref{w01} one can easily check that the second relation in
\eqref{u1a} is automatically satsfied
\beq
\overline{W}_x(n)\,\overline{W}_{-\qs/x}(n)=1\,.
\eeq
Similar considerations apply to the
weight function $W_x(n)$. The results are presented in
\eqref{weights-V} of the main text.

\app{Factorization of the partition function\label{appB}} 

In Sect.\ref{facpar} we have considered an example of the planar graph
${\mathscr L}$, where one spectral parameter line (the leftmost line in
Fig.~\ref{fig-net5}) could be completely disentangled from the rest of
the graph by using the $Z$-invariance.  
Here we present a more detailed description of yet another example of
this sort.

The definition of the $Z$-invariant models formulated in 
Sect.~\ref{ordef} above can be generalized by
including additional types of lines. For instance, the
six-vertex type lines carring two-state spins, which have already appeared
in Sect.~\ref{face-vert}. 
Graphically, they are represented as thick directed lines, which are continued
as dashed lines in the shaded areas. 
Their intersections with the thin spectral parameter lines together
with the associated tri-spin weights 
are shown in \eqref{thick-vert} and \eqref{thick-vert-2} (the weights
are defined in \eqref{Phi-def}, \eqref{Omega-def}).

\bigskip
\begin{figure}[ht]
\begin{center}
\raisebox{-.5\height}{\includegraphics[scale=.89]{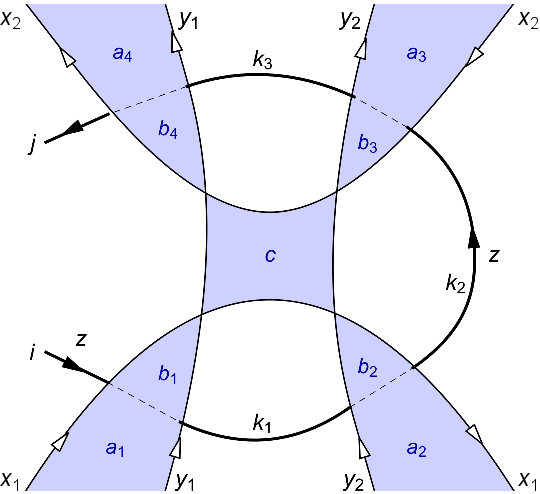}}
$\Longrightarrow$\raisebox{-.5\height}{\includegraphics[scale=0.89]
{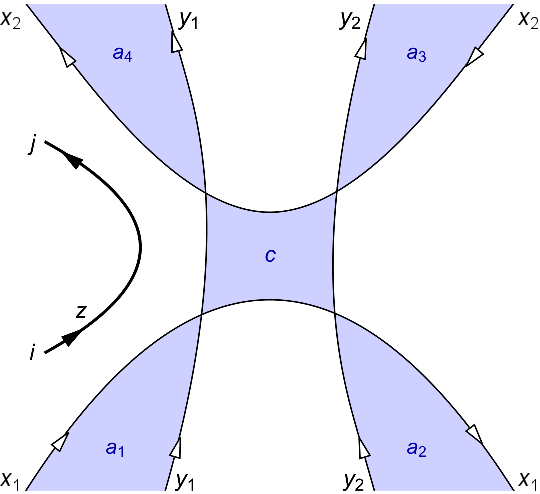}}
\end{center}
\caption{An example of planar graphs connected with each other by the
Yang-Baxter type and inversion relation moves. Their partition
functions differ by  
simple factors, see \eqref{Z-rhs} }
\label{fig-net10}
\end{figure}
Apart from adding the new type of lines, we will also slightly relax the
requirements on the topology of the graph ${\mathscr L}$ discussed in 
Sect.~\ref{ordef}. This graph is formed by a 
set of directed spectral parameter lines, which go from the bottom of
the graph to the top, intersecting one another on the way. To formalize
this one can say that the lines start at the bottom side of a rectangular 
strip and end up at its top side. 
Let us now relax this by allowing 
the lines (i) to start from the bottom or the lateral sides of the strip, (ii) 
to end up at at the top or the lateral sides of the strip. However,
as before, there should be no closed directed paths. As an example,
consider the graph presented on the left side of
Fig.~\ref{fig-net10} (we assume that the line $x_1$ is ending at the right
side of the strip, while the line
$x_2$ is starting from it).    
It is easy to check that there are no closed directed loops, so that
the graph satisfy all the above conditions.

Next, we use the graphic rules \eqref{thick-vert}, \eqref{thick-vert-2} 
and \eqref{W-shade}  to assign Boltzmann weights to all vertices
in Fig.~\ref{fig-net10}. The partition function is defined as a sum
over all configurations of interior spins (in this case
$b_1,b_2,b_3,b_4,c\in {\mathbb Z}$ and $k_1,k_2,k_3=\pm1$)
 with the
weight equal to the product of the local weights over all
vertices. Note, that the vertex weights only depend on the relative
orientation of the directed lines among themselves and with respect to
the shaded faces. They are not affected by overall rotations of 
the vertices. Therefore, the partition function of the graph will not
change, if the graph is deformed without changing its topology. 
\bigskip
\begin{figure}[ht]
\begin{center}
\includegraphics[scale=1.4]{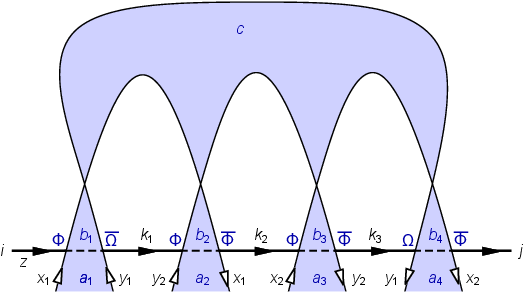}
\end{center}
\caption{An equivalent transformation of the graph shown on the left
side of Fig.~\ref{fig-net10}. Note, that all lines in the new graph  
are heading in
the same direction: from left to right.
}
\label{fig-net11}
\end{figure}
In this way the graph on the left side of Fig.~\ref{fig-net10} can be 
transformed into an equivalent graph shown in
Fig.~\ref{fig-net11}. Its partition function reads
\be\label{Z-app}
{\cal Z}=\sum_{b_1,b_2,b_3,b_4\in{\mathbb Z}}\,\big({\mathcal M}(x_1,x_2,y_1,y_2\,|z)_{a_1,a_2,a_3,a_4}^{\,b_1,b_2,b_3,b_4}\big)_{i,j} \  {\cal V}(x_1,x_2,y_1,y_2)_{b_1,b_2,b_3,b_4}
\ee
where
\be\label{V-app}
{\cal V}(x_1,x_2,y_1,y_2)_{b_1,b_2,b_3,b_4}=\sum_{c\in{\mathbb Z}}\,
\overline{W}_{x_1/y_1}(b_1-c)\,W_{x_1/y_2}(b_2-c)\,W_{y_2/x_2}(b_3-c)\,
\overline{W}_{y_1/x_2}(b_4-c)\,,
\ee
combines contributions of the two-spin weights, as defined in \eqref{W-shade} (top part of 
Fig.~\ref{fig-net11}) and 
\be\label{M-def}
\begin{array}{rcl}
\ds\big({\mathcal M}
(x_1,x_2,y_1,y_2\,|z)_{a_1,a_2,a_3,a_4}^{\,b_1,b_2,b_3,b_4}\big)_{i,j}
\ds&=&\ds
\sum_{k_1,k_2,k_3=\pm1}\,\Phi(x_1,z)_{a_1,i}^{b_1}
\,{\overline\Omega}(y_1,z)_{a_1,k_1}^{b_1}
\,\Phi(y_2,z)_{a_2,k_1}^{b_2}
\,{\overline\Phi}(x_1,z)_{a_2,k_2}^{b_2}\,
\times\qquad\qquad\qquad\\[.45cm]
&&\qquad\qquad\times\,
\Phi(x_2,z)_{a_3,k_2}^{b_3}
\,{\overline\Phi}(y_2,z)_{a_3,k_3}^{b_3}
\,\Omega(y_1,z)_{a_4,k_3}^{b_4}\,{\overline\Phi}(x_2,z)_{a_4,j}^{b_4}
{}_{\phantom{\Vert}}\,,
\end{array}
\ee
combines the tri-spin weights, graphically defined
in \eqref{thick-vert}, \eqref{thick-vert-2} (bottom part of
Fig.~\ref{fig-net11}). Note, that the last expression could be viewed
as matrix elements of a (column-inhomogeneous) monodromy matrix 
\be\label{M-def2}
\begin{array}{l}
\ds\big({\mathcal M}
(x_1,x_2,y_1,y_2\,|z)_{a_1,a_2,a_3,a_4}^{\,b_1,b_2,b_3,b_4}\big)_{i,j}
\ds=\ds\\[.5cm]
\ds
\ds\qquad=\sum_{k_1,k_2,k_3}\big({\mathcal L}(x_1,y_1,z)_{a_1}^{b_1}\big)_{i,k_1}\,
\big({\mathcal L}(y_2,x_1/\xi,z)_{a_2}^{b_2}\big)_{k_1,k_2}\,
\big({\mathcal L}(x_2,y_2/\xi,z)_{a_3}^{b_3}\big)_{k_2,k_3}\,
\big({\mathcal L}(y_1/\xi,x_2/\xi,z)_{a_4}^{b_4}\big)_{k_3,j}\,.
\end{array}
\ee
with $i,j=\pm1$. In writing the last formula 
we have used \eqref{Omega-def} and \eqref{LL2}. It is a
two-by-two matrix $\bm{\mathcal M}(x_1,x_2,y_1,y_2\,|z)$ with
operator-valued elements acting the ``quantum space'' ${\mathbb
Z}^4$. From \eqref{RLL} it follows that 
\beq\label{RMM}
\sum_{j_1,j_2=\pm1}\;\bm{{\mathcal M}}(z_1)_{i_1,j_1}\,
\bm{{\mathcal M}}(z_2)_{i_2,j_2}\,
\Rc^{\rm (6v)}\big(z_1^2/z_2^2\big)_{j_1,j_2}^{k_1,k_2}
=
\sum_{j_1,j_2=\pm1}\;
\Rc^{\rm (6v)}\big(z_1^2/z_2^2\big)_{i_1,i_2}^{j_1,j_2}\, 
\bm{{\mathcal M}}(z_2)_{j_2,k_2}\,\bm{{\mathcal M}}(z_1)_{j_1,k_1}\;,
\eeq
where we have omitted the arguments $\{x_1,x_2,y_1,y_2\}$, which are the
same for $\bm{{\mathcal M}}(z_1)$ and
$\bm{{\mathcal M}}(z_2)$. 

\begin{figure}[ht]
\begin{center}
\includegraphics[scale=1.3]{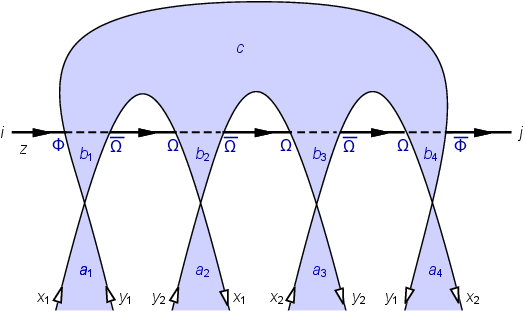}
\end{center}
\caption{The graph obtained from Fig.~\ref{fig-net11} by four Yang-Baxter type moves, shown in \eqref{peq2a}, \eqref{peq3} and \eqref{peq4}. }
\label{fig-net12}
\end{figure}

Consider now further equivalence 
transformations of the graph in Fig.\ref{fig-net11} by moving
the thick horizontal $z$-line upwards through the intersections of the
thin lines to the position shown in Fig.~\ref{fig-net12}. In doing 
\bigskip
this we need to use three similar, but slightly different, Yang-Baxter
type moves. The first one is based on the relation \eqref{psieq2},
which for readers' convenience we reproduce here
\beq
\label{psieq2a}
\sum_{b} \Phi(x,z)_{a,i}^{b} \ 
\overline{\Omega}(y,z)_{a,i'}^{b}\  
\overline{W}_{x/y}(b-c)
\;=\;
\sum_{b} \overline{W}_{x/y}(a-b)\  \Phi(y,z)_{b,i}^{c} \  
\overline{\Omega}(x,z)_{b,i'}^{c}\;.
\eeq
It can be represented graphically as  
%
\begin{equation}\label{peq2a}
\begin{tikzpicture}[scale=0.6,baseline=(current  bounding  box.center)]
\draw [fill, opacity=0.35, blue,path fading=north] (0,2) -- (-1,4) -- (0,4.5) -- (1,4) -- (0,2);
\draw [fill, opacity=0.3, blue,path fading=south] (0,2) -- (1.5,-1) -- (0,-1.5) -- (-1.5,-1) -- (0,2);
\draw [-open triangle 45, black, thin] (-1.5,-1) -- (1,4); \node [above] at (1,4) {$x$};
\draw [-open triangle 45, black, thin] (1.5,-1) -- (-1,4); \node [above] at (-1,4) {$y$};
\draw [dashed, thick] (1,0) -- (0,0); \draw [dashed, thick] (0,0) -- (-1,0);
\draw[-latex, thick] (-2,0) -- (-1.3,0); \draw [thick] (-1.,0) -- (-2,0);  
\draw[-latex, thick] (1,0) -- (1.7,0); \draw [thick] (2.,0) -- (1,0); 
\node [left] at (-2,0) {$i$}; \node [right] at (2,0) {$i'$};
\node [below] at (-1.75,0) {$z$};
\node [below] at (1.6,0) {$z$};
\node at (0,0.66) {$b$}; \node[above] at (0,3.35) {$c$};
\node [below] at (0,-0.3) {$a$};
\node [anchor=south east] at (-1,0) {$\Phi$};
\node [anchor=south west] at (1,0) {$\overline{\Omega}$};
\node [right] at (3.5,1.5) {$=$};
\end{tikzpicture}
\hskip 10mm
\begin{tikzpicture}[scale=0.6,baseline=(current  bounding  box.center)]
\draw [fill, opacity=0.3, blue, path fading=north] (0,1) -- (-1.5,4) -- (0,4.5) -- (1.5,4) -- (0,1);
\draw [fill, opacity=0.35, blue,path fading=south] (0,1) -- (1,-1) -- (0,-1.5) -- (-1,-1) -- (0,1);
\draw [-open triangle 45, black, thin] (-1,-1) -- (1.5,4); \node [above] at (1.5,4) {$x$};
\draw [-open triangle 45, black, thin] (1,-1) -- (-1.5,4); \node [above] at (-1.5,4) {$y$};
\draw [dashed, thick] (1,3) -- (0,3); \draw [dashed, thick] (0,3) -- (-1,3);
\draw[-latex, thick] (-2,3) -- (-1.3,3); \draw [thick] (-1.,3) -- (-2,3);  
\draw[-latex, thick] (1,3) -- (1.7,3); \draw [thick] (2.,3) -- (1,3); 
\node [left] at (-2,3) {$i$}; \node [right] at (2,3) {$i'$};
\node [above] at (-1.75,3) {$z$};
\node [above] at (1.6,3) {$z$};
\node at (0,-0.35) {$a$}; \node at (0,2.34) {$b$};
\node [above] at (0,3.35) {$c$};
\node [anchor=north east] at (-1,3) {$\Phi$};
\node [anchor=north west] at (1,3) {$\overline{\Omega}$};
\end{tikzpicture}
\end{equation} 
It is worth noting that \eqref{psieq2a} can be rewritten as 
\be
\sum_b\,(\bm{\mathcal L}(x,y,z))_a^b\, \overline{W}_{x/y}(b-c)
=\sum_b\,\overline{W}_{x/y}(a-b)\,(\bm{\mathcal L}(y,x,z))_b^c\,.
\ee
The other two required relations are simple corollaries of \eqref{psieq2a}, 
namely,  
\be\label{psieq3}
\sum_b\,\Phi(x,z)_{a,i}^b\,\overline{\Phi}(y,z)_{a,i'}^b\,W_{y/x}(b-c)=
\sum_b\,W_{y/x}(a-b)\,
\Omega(y,z)_{b,i}^c\,\overline{\Omega}(x,z)_{b,i'}^c\,,
\ee
which is represented as 
%
\begin{equation}\label{peq3}
\begin{tikzpicture}[scale=0.6,baseline=(current  bounding  box.center)]
\draw [fill, opacity=0.3, blue,path fading=north] (0,2) -- (-1,4) -- (0,4.5) -- (1,4) -- (0,2);
\draw [fill, opacity=0.25, blue,path fading=south] (0,2) -- (1.5,-1) -- (0,-1.5) -- (-1.5,-1) -- (0,2);
\draw [-open triangle 45, black, thin] (-1.5,-1) -- (1,4); \node [above] at (1,4) {$x$};
\draw [-open triangle 45, black, thin] (-1,4)-- (1.5,-1); \node [above] at (-1,4) {$y$};
\draw [dashed, thick] (1,0) -- (0,0); \draw [dashed, thick] (0,0) -- (-1,0);
\draw[-latex, thick] (-2,0) -- (-1.3,0); \draw [thick] (-1.,0) -- (-2,0);  
\draw[-latex, thick] (1,0) -- (1.7,0); \draw [thick] (2.,0) -- (1,0); 
\node [left] at (-2,0) {$i$}; \node [right] at (2,0) {$i'$};
\node [below] at (-1.75,0) {$z$};
\node [below] at (1.8,0) {$z$};
\node at (0,0.66) {$b$}; \node[above] at (0,3.35) {$c$};
\node [below] at (0,-0.3) {$a$};
\node [anchor=south east] at (-1,0) {$\Phi$};
\node [anchor=south west] at (1,0) {$\overline{\Phi}$};
\node [right] at (3.5,1.5) {$=$};
\end{tikzpicture}
\hskip 10mm
\begin{tikzpicture}[scale=0.6,baseline=(current  bounding  box.center)]
\draw [fill, opacity=0.25, blue, path fading=north] (0,1) -- (-1.5,4) -- (0,4.5) -- (1.5,4) -- (0,1);
\draw [fill, opacity=0.3, blue,path fading=south] (0,1) -- (1,-1) -- (0,-1.5) -- (-1,-1) -- (0,1);
\draw [-open triangle 45, black, thin] (-1,-1) -- (1.5,4); \node [above] at (1.5,4) {$x$};
\draw [-open triangle 45, black, thin]  (-1.5,4)--(1,-1); \node [above] at (-1.5,4) {$y$};
\draw [dashed, thick] (1,3) -- (0,3); \draw [dashed, thick] (0,3) -- (-1,3);
\draw[-latex, thick] (-2,3) -- (-1.3,3); \draw [thick] (-1.,3) -- (-2,3);  
\draw[-latex, thick] (1,3) -- (1.7,3); \draw [thick] (2.,3) -- (1,3); 
\node [left] at (-2,3) {$i$}; \node [right] at (2,3) {$i'$};
\node [above] at (-1.75,3) {$z$};
\node [above] at (1.8,3) {$z$};
\node at (0,-0.35) {$a$}; \node at (0,2.34) {$b$};
\node [above] at (0,3.35) {$c$};
\node [anchor=north east] at (-1,3) {$\Omega$};
\node [anchor=north west] at (1,3) {$\overline{\Omega}$};
\end{tikzpicture}
\end{equation}
and 
\be\label{psieq4}
\sum_b\,\Omega(x,z)_{a,i}^b\,\overline{\Phi}(y,z)_{a,i'}^b\, 
\overline{W}_{x/y}(b-c)=     
\sum_b\,\overline{W}_{x/y}(a-b)\,
\Omega(y,z)_{b,i}^c\,\overline{\Phi}(x,z)_{b,i'}^c\,,
\ee
represented as
%
\begin{equation}\label{peq4}
\begin{tikzpicture}[scale=0.6,baseline=(current  bounding  box.center)]
\draw [fill, opacity=0.3, blue,path fading=north] (0,2) -- (-1,4) -- (0,4.5) -- (1,4) -- (0,2);
\draw [fill, opacity=0.25, blue,path fading=south] (0,2) -- (1.5,-1) -- (0,-1.5) -- (-1.5,-1) -- (0,2);
\draw [-open triangle 45, black, thin]  (1,4)--(-1.5,-1); 
\node [above] at (1,4) {$x$};
\draw [-open triangle 45, black, thin] (-1,4)-- (1.5,-1); \node [above] at (-1,4) {$y$};
\draw [dashed, thick] (1,0) -- (0,0); \draw [dashed, thick] (0,0) -- (-1,0);
\draw[-latex, thick] (-2,0) -- (-1.3,0); \draw [thick] (-1.,0) -- (-2,0);  
\draw[-latex, thick] (1,0) -- (1.7,0); \draw [thick] (2.,0) -- (1,0); 
\node [left] at (-2,0) {$i$}; \node [right] at (2,0) {$i'$};
\node [below] at (-1.75,0) {$z$};
\node [below] at (1.8,0) {$z$};
\node at (0,0.66) {$b$}; \node[above] at (0,3.35) {$c$};
\node [below] at (0,-0.3) {$a$};
\node [anchor=south east] at (-1,0) {$\Omega$};
\node [anchor=south west] at (1,0) {$\overline{\Phi}$};
\node [right] at (3.5,1.5) {$=$};
\end{tikzpicture}
\hskip 10mm
\begin{tikzpicture}[scale=0.6,baseline=(current  bounding  box.center)]
\draw [fill, opacity=0.25, blue, path fading=north] (0,1) -- (-1.5,4) -- (0,4.5) -- (1.5,4) -- (0,1);
\draw [fill, opacity=0.3, blue,path fading=south] (0,1) -- (1,-1) -- (0,-1.5) -- (-1,-1) -- (0,1);
\draw [-open triangle 45, black, thin] (1.5,4)--(-1,-1); \node [above] at (1.5,4) {$x$};
\draw [-open triangle 45, black, thin]  (-1.5,4)--(1,-1); \node [above] at (-1.5,4) {$y$};
\draw [dashed, thick] (1,3) -- (0,3); \draw [dashed, thick] (0,3) -- (-1,3);
\draw[-latex, thick] (-2,3) -- (-1.3,3); \draw [thick] (-1.,3) -- (-2,3);  
\draw[-latex, thick] (1,3) -- (1.7,3); \draw [thick] (2.,3) -- (1,3); 
\node [left] at (-2,3) {$i$}; \node [right] at (2,3) {$i'$};
\node [above] at (-1.75,3) {$z$};
\node [above] at (1.8,3) {$z$};
\node at (0,-0.35) {$a$}; \node at (0,2.34) {$b$};
\node [above] at (0,3.35) {$c$};
\node [anchor=north east] at (-1,3) {$\Omega$};
\node [anchor=north west] at (1,3) {$\overline{\Phi}$};
\end{tikzpicture}
\end{equation}
Taking into account \eqref{Omega-def} and \eqref{crossing2} it is easy
to see that \eqref{psieq3} follows from \eqref{psieq2a} with the substitution
$y\to y/\xi$. Similarly, \eqref{psieq4} 
is obtained from \eqref{psieq2a} with $x\to x/\xi$ and $y\to y/\xi$.

Next, we continue moving the $z$-line in Fig.~\ref{fig-net12} further
upwards to completely detach it from the rest of the graph. To do this
we will use the unversion relations (see \eqref{Phi-ort}
and \eqref{omega-ort} of the main text)
\be\label{unit00}
\sum_{i=\pm1} \overline{\Omega}(x,z)_{a,i}^{a'}\,
\Omega(x,z)_{b,i}^{a'}\;=\;[x^2/z^2]\,\delta_{a,b}\;,
\qquad
\ds\sum_{a'\in{\mathbb Z}} \Phi(x,z)_{a,i}^{a'}\,
\overline{\Phi}(x,z)_{a,i'}^{a'}
\;=\;[x^2/z^2]\,
\delta_{i,i'}\;.
\ee
with the notation $[x]=x-1/x$.
The first of them is represented as 
\bigskip
\be\label{unit10}
\begin{tikzpicture}[scale=.9,baseline={(0,0)}]
\draw[fill, opacity=0.4,color={rgb:red,.05;green,.05;blue,.5},path fading=north]
(-1.2,-1.5) -- (0,-1.5) -- (0,0) .. controls (0,1.5) and (1.5,1.5) .. (1.5,0) -- 
(1.5,-1.5) -- (2.7,-1.5) -- (2.7,2) -- (-1.2,2) ;
\draw [dashed,thick] (-1.2,0) -- (2.7,0);
\draw [-open triangle 45, thin,opacity=1] (1.5,0) -- (1.5,-1.3);
\draw [thin,opacity=1] (1.5,-1.5) -- (1.5,-1.3);
\node [left] at (1.5,-1.7) {$x$};
\draw [-open triangle 45, thin,opacity=1] (0,-1.5) -- (0,-1); 
\node [right] at (0,-1.7) {$x$};
\draw [thin] (0,-1.5) -- (0,0); 
\draw[thin]
(0,-1.5) -- (0,0) .. controls (0,1.5) and (1.5,1.5) .. (1.5,0) 
-- (1.5,-1.5);
\node [anchor=north west] at (0,0) {$\overline{\Omega}$};
\node [anchor=north east] at (1.5,-.08) {$\Omega$};
\node at (.7,1.5) {$a'$}; 
\node [left] at (-.3,-.8) {$a$}; \node [right] at (1.8,-.8) {$b$};
\node [above] at (.2,0) {$z$};
\node [above] at (1.2,0) {$i$};
\draw [-latex, thick] (0,0) -- (.75,0); \draw [thick] (0,0) -- (1.5,0);
\end{tikzpicture}
\begin{tikzpicture}[scale=.9,baseline={(0,0)}]
\node at (-2.6,0) {$\;\;=\;\,[x^2/z^2]\,\;$}; 
\draw[fill, opacity=0.4,color={rgb:red,.05;green,.05;blue,.5},path fading=north]
(-1.2,-1.5) -- (0,-1.5) -- (0,-1) .. controls (0,-.0) and (1.5,-.0) .. (1.5,-1) -- 
(1.5,-1.5) -- (2.7,-1.5) -- (2.7,2) -- (-1.2,2) ;
\draw [dashed,thick] (-1.2,0) -- (2.7,0);
\draw[thin]
(0,-1) .. controls (0,-.0) and (1.5,-.0) .. (1.5,-1) ;
\node at (.7,1.) {$a'$}; 
\node [left] at (-.3,-.8) {$a$}; \node [right] at (1.8,-.8) {$b$};
\draw [-open triangle 45, thin,opacity=1] (1.5,-1) -- (1.5,-1.3);
\draw [thin,opacity=1] (1.5,-1.5) -- (1.5,-1.3);
\node [left] at (1.5,-1.7) {$x$};
\draw [-{open triangle 45}, thin,opacity=1] (0,-1.5) -- (0,-1); 
\node [right] at (0,-1.7) {$x$};
\end{tikzpicture}
\ee
and the second one as
\begin{equation}\label{unit20}
\begin{tikzpicture}[scale=1.2,baseline=(current  bounding  box.center)]
\draw [fill, opacity=0.3,color={rgb:red,.05;green,.05;blue,.5},path fading=south] (1.5,-1.5) -- (1.5,0) --  (2.5,0) -- (2.5,-1.5) -- (2,-1.7) -- (1.5,-1.5);
\draw [-open triangle 45, thin,opacity=1] (1.5,-1.5) -- (1.5,-1); \node [left] at (1.5,-1.3) {$x^{\phantom{'}}$};
\draw [thin] (1.5,-1.5) -- (1.5,-1.3);
\draw [thin] (1.5,-1.) -- (1.5,0); 
\draw [thin] (2.5,-1.3) -- (2.5,-1.5); 
\draw [-open triangle 45, thin,opacity=1] (2.5,0) -- (2.5,-1.3); \node [right] at (2.5,-1.3) {$x$};
\draw [thick] (1.5,0) -- (0.8,0); 
\draw [-latex, thick] (0,0) -- (0.8,0);
\node [left] at (0,0) {$i$}; \node [right] at (4,0) {$i'$};
\node [below] at (0.45,0) {$z$};
\node [above] at (3.3,0) {$z$};
\draw [thick, dashed] (2.5,0) -- (1.5,0); 
\draw [thick] (4,0) -- (3.5,0); 
\draw [-latex,thick] (2.5,0) -- (3.5,0);

\node [anchor=south east] at (1.5,0) {$\Phi$};
\node [anchor=north west] at (2.5,0) {$\overline\Phi$};
\node [below] at (2,-.5) {$a$};
\node [above] at (2,.3) {$a'$};
\draw [fill, opacity=0.3,color={rgb:red,.05;green,.05;blue,.5}]
(1.5,0) .. controls (1.5,1.5) and (2.5,1.5) .. (2.5,0);
\draw [thin]
(1.5,0) .. controls (1.5,1.5) and (2.5,1.5) .. (2.5,0);

\node [right] at (4.4,0) {$\,=\,\,\,[x^2/z^2]$};
\begin{scope}[xshift=.6cm]
\draw [-latex,thick] (6,0) -- (7,0);
\draw [thick] (7,0) -- (8,0);
\node [left] at (6,0) {$i$};
\node [right] at (8,0) {$i'$};
\draw [fill, opacity=0.4,color={rgb:red,.05;green,.05;blue,.5},path fading=south]
(6.5,-1.5) -- (6.5,-1) .. controls (6.5,0.1) and (7.5,0.1) .. (7.5,-1) --
(7.5,-1.5)--(7,-1.7);
\draw [thin]
(6.5,-1.5) -- (6.5,-1) .. controls (6.5,0.1) and (7.5,0.1) .. (7.5,-1) --
(7.5,-1.5);
\draw [-open triangle 45, thin,opacity=1] (6.5,-1.5) -- (6.5,-1); \node [left] at (6.5,-1.3) {$x^{\phantom{'}}$};
\draw [-open triangle 45, thin,opacity=1] (7.5,-1) -- (7.5,-1.3); \node [right] at (7.5,-1.3) {$x$};
\node [below] at (7,-.5) {$a$};
\node [above] at (7,0) {$z$};
\end{scope}
\end{tikzpicture}
\end{equation}
Proceeding in this way one obtains for the partition
function \eqref{Z-app},
\be\label{Z-rhs}
\begin{array}{l}
\ds{\mathcal Z}=\sum_{\{b\}} \,
\big(\,{{\mathcal M}}(x_1,x_2,y_1,y_2\,|z_2)_{\{a\}}^{\{b\}}\,\big)_{i,j}\ 
{\cal V}(x_1,x_2,y_1,y_2)_{\{b\}}\\[.4cm]
\qquad\qquad\qquad\qquad\qquad
=\ds\delta_{i,j}\,\,[x_1^2/z^2]\ 
[x_2^2/z^2]\,[y_1^2/z^2]\,[y_2^2/z^2]\ \ 
{\cal V}(x_1,x_2,y_1,y_2)_{\{a\}}
\end{array}
\ee
where $\{a\}=\{a_1,a_2,a_3,a_4\}$ and similarly for $\{b\}$.
It is not difficult to see that up to the scalar factor the RHS of the
last relation precisely reduces to the partition function of the graph
shown on the right side of Fig.~\ref{fig-net10}. Obviously, the
quantity \eqref{V-app} can be regarded as an eigenvector of the
monodromy matrix \eqref{M-def}. Similar results for the Yangian case were  
obtained in \cite{Chicherin:2017cns}.

\app{The 8-vertex free-fermion model\label{appC}} 
Note, that when 
\beq\label{qeps1}
\qs=-\re^{-\epsilon}\,,\qquad \epsilon\to 0\,,\qquad \epsilon>0\,,
\eeq
the 8-vertex model becomes
critical \cite{Baxter:1982zz}. Moreover,
in our case \eqref{xi-square} we also have the relation  
\beq\label{qeps2}
\overline{\xi}=\re^{\ri \lambda}\,,\qquad \lambda=\half\ri \,\epsilon\,.
\eeq
To study this point it is convenient to
express the weights \eqref{abcd-8v} in terms of the theta functions
\beq
\Theta_i(v)=\vartheta_i(v,\widetilde{\qs})\,,
\eeq 
of the nome $\widetilde{\qs}$ and the variable $v$,
\beq
\widetilde{\qs}=\re^{-{\pi^2}/{\epsilon}}\,,\qquad
v=-\ri\,\pi u /\epsilon\,.
\eeq
Using the standard formulae for transformations of the elliptic
functions for the weights \eqref{abcd-8v},
one obtains,
\beq\label{abcd-8v2}
\begin{array}{rclrcl}
\as&=&\rho'_8(v)\,\Theta_4\,\Theta_2(v)\,\Theta_3(v)\,,
\qquad\qquad\qquad&  
\bs&=&\;\rho'_8(v)\,\Theta_4\,\Theta_1(v)\,\Theta_4(v)\,,\\[.3cm]
\cs&=&\rho'_8(v)\,\Theta_2\,\Theta_3(v)\,\Theta_4(v)\,,\qquad&
\dds&=&\ri\,\rho'_8(v)\,\Theta_2\,\Theta_1(v)\,\Theta_2(v)\,.
\end{array}
\eeq
where
\beq
\rho'_8(v)=\rho_8\,\Big(\frac{\epsilon}{\pi}\Big)^{-3/2}\,
\exp\Big\{+\frac{3\ri \pi}{4}
+\frac{\epsilon}{2}  -\frac{2\epsilon\,v (\pi/2-v)}{\pi^2}\Big\}
\eeq
In the leading order at $\epsilon\to0$, one obtains
\beq\label{abcd-6v}
\begin{array}{rclrcl}
\as&=&\rho''_8\,\cos v\,,
\qquad\qquad\qquad&  
\bs&=&\;\rho''_8\,\sin v\,,\\[.3cm]
\cs&=&\rho''_8\,,\qquad&
\dds&=&0\,,
\end{array}
\eeq
which is the six-vertex free-fermion model (the ratio $\dds/\cs$ in this
limit vanishes as $\exp(-\pi^2/2\epsilon)$).  
Here
\beq
\rho''_8=2\rho_8\,\Big(\frac{\epsilon}{\pi}\Big)^{-3/2}
\,\exp\Big\{-\frac{\pi^2}{4\epsilon}+\frac{3\ri\pi}{4}\Big\}\,,
\eeq
With \eqref{abcd-6v} the partition
function  \eqref{kappa-ffm} becomes
\bea\nonumber
\log\kappa^{(8)}&=&\log\rho''_8
+\frac{1}{8 \pi^2}
\iint_0^{2\pi}\rd\phi_1 \rd\phi_2\,
\log\Big(2+2\sin^2v\,\cos(\phi_1-\phi_2)-2 
\cos^2v\,\cos(\phi_1+\phi_2)\Big)+O(\epsilon)\\[.3cm]
&=&\log\rho''_8+\frac{1}{4 \pi}\int_0^{2\pi}
\rd\phi\log\Big(1+\sin(2 v)\,|\cos\phi\,|\Big)+O(\epsilon)\,,
\eea
Evaluating the integral and replacing the variable $v$ by a new
 variable  $\theta=2v/\pi$,
 one obtains 
\beq
\kappa^{(8)}= \frac{\rho''_8\,G^2(\theta)\,G^2(1-\theta)}{G^2(1)}\,\big(1+O(\epsilon)\big)\,,
\qquad \qquad \theta=\frac{2v} {\pi}\,,
\eeq
where
\beq
G(\theta)=\exp
\left\{\frac{1}{\pi}\int_0^{\pi \theta/2} (x\cot x)\,\rd x\,\right\}=
\re^{\frac{1}{2}\theta}\,\prod_{n=1}^{\infty}\left(
\frac{\Gamma\big(n+\frac{\theta}{2}\big)}{
\Gamma\big(n-\frac{\theta}{2}\big)}\re^{-\theta \psi(n)}\right)
\eeq
and $\psi(x)=\rd \log\Gamma(x)/\rd x$ is the logarithmic derivative of
the gamma-function. On the other hand, the same result should, 
of course, follow
from the general expression of the partition function of the 8-vertex
model \eqref{kappa-8v} or from the corresponding expression for the
six-vertex model \eqref{abcd-6v} (with $v=\pi\theta/2$). Indeed, using
Eq.(8.8.17) from \cite{Baxter:1982zz}, one obtains 
\beq
\kappa^{(8)}=\rho''_8\,\cos({\pi\theta}/{2})\,
\exp\left(\frac{1}{4}\int_{\mathbb{R}} \frac{dw}{w} \;\frac{\sinh(2\theta
w)}{\cosh^2(w)}\right)\big(1+O(\epsilon)\big)\,,\qquad \epsilon\to0\,.
\eeq


\begin{thebibliography}{99}
\addcontentsline{toc}{section}{Bibliography.} 
\bibitem{Baxter:1982zz}
R.~J.~Baxter,
``Exactly solved models in statistical mechanics,''
\href{https://physics.anu.edu.au/research/ftp/mpg/baxter_book.php}{Academic: London, 1982.}

\bibitem{Onsager:1944}
L.~Onsager, 
\newblock ``Crystal statistics. {I}. {A} two-dimensional model with an
  order-disorder transition,''
\newblock 
\href{https://doi.org/10.1103/PhysRev.65.117}{Phys. Rev. {\bf 65} (1944) 117--149}.

\bibitem{McGuire:1964}
J.~B.~McGuire, ``Study of exactly solvable one-dimensional $N$-body
problems,'' 
\href{https://doi.org/10.1063/1.1704156}{J. Math. Phys. {\bf 5} (1964) 622--636}.

\bibitem{Yang:1967}
C.~N.~Yang, 
\newblock ``Some exact results for the many-body problem in one dimension with
  repulsive delta-function interaction,''
\newblock 
\href{https://doi.org/10.1103/PhysRevLett.19.1312}{Phys. Rev. Lett. {\bf 19} (1967) 1312--1315}.

\bibitem{Baxter:1972}
R.~J.~Baxter, 
\newblock ``Partition function of the eight-vertex lattice model,''
\newblock 
\href{https://doi.org/10.1016/0003-4916(72)90335-1}{Ann. Physics {\bf 70} (1972) 193--228}.

\bibitem{Kashiwara:1986}
M.~Kashiwara, and T.~Miwa,
\newblock A class of elliptic solutions to the star-triangle relation.
\newblock \href{https://doi.org/10.1016/0550-3213(86)90591-2}{Nucl. Phys. B {\bf 275} (1986) 121--134.}

\bibitem{vG85}
G.~von Gehlen  and V.~Rittenberg,
\newblock $Z(n)$-symmetric quantum chains with an infinite set of conserved
  charges and $Z(n)$ zero modes.
\newblock \href{https://doi.org/10.1016/0550-3213(85)90350-5}{Nucl. Phys. B {\bf 257} (1985) 351.}

\bibitem{Au-Yang:1987syu}
H.~Au-Yang, B.~M.~McCoy, J.~H.~H.~Perk, S.~Tang and M.~L.~Yan,
\newblock ``Commuting transfer matrices in the chiral Potts models: Solutions of Star triangle equations with genus $>$ 1'',
\newblock \href{https://doi.org/10.1016/0375-9601(87)90065-X}
{Phys. Lett.  A \textbf{123} (1987)  219--223}.


\bibitem{Baxter:1987eq}
R.~J.~Baxter, J.~H.~H.~Perk and H.~Au-Yang,
\newblock ``New solutions of the star triangle relations for the chiral Potts model'',
\newblock \href{https://doi.org/10.1016/0375-9601(88)90896-1}
{Phys. Lett.  A \textbf{128} (1988) 138--142}.

\bibitem{FZ82}
V.~A.~Fateev and A.~B.~Zamolodchikov, 
\newblock Self-dual solutions of the star-triangle relations in
  {$Z_{N}$}-models.
\newblock \href{https://doi.org/10.1016/0375-9601(82)90736-8}{Phys. Lett. A {\bf 92} (1982) 37--39.}




\bibitem{Bax02rip}
R.~J.~Baxter, 
\newblock A rapidity-independent parameter in the star-triangle relation.
\newblock In {\em Math{P}hys {O}dyssey, 2001}, {\em Prog. Math.
  Phys.}, {\bf 23}, 9--63. Birkh\"auser Boston, Boston, MA, 2002,
\newblock \href{https://arxiv.org/abs/cond-mat/0108363}{\texttt{	arXiv:cond-mat/0108363}}.


\bibitem{Zam-fish}
A.~B.~Zamolodchikov, 
\newblock ``{F}ishing-net'' diagrams as a completely integrable system.
\newblock \href{https://doi.org/10.1016/0370-2693(80)90547-X}{Phys. Lett. B {\bf 97} (1980) 63--66.}

\bibitem{FV95}
A.~Y.~Volkov, and L.~D~ Faddeev, 
\newblock Yang-{B}axterization of the quantum dilogarithm.
\newblock Zapiski Nauchnykh Seminarov POMI {\bf 224} (1995) 146--154.
\newblock English translation: \href{https://link.springer.com/article/10.1007/BF02364981}{J. Math. Sci. {\bf 88} (1998) 202-207.}

\bibitem{BMS07a}
V.~V.~Bazhanov, V.~V.~ Mangazeev, and S.~M.~Sergeev, 
\newblock Faddeev-Volkov solution of the Yang-Baxter Equation and Discrete
  Conformal Symmetry.
\newblock \href{https://doi.org/10.1016/j.nuclphysb.2007.05.013}{Nucl. Phys. B {\bf 784} (2007) 234--258},
\newblock \href{https://arxiv.org/abs/hep-th/0703041}{\texttt{arXiv:hep-th/0703041}}.


\bibitem{BSp}
A.~I.~Bobenko, and B.~A.~Springborn, 
\newblock Variational principles for circle patterns and Koebe's theorem.
\newblock \href{https://www.ams.org/journals/tran/2004-356-02/S0002-9947-03-03239-2/S0002-9947-03-03239-2.pdf}{Trans. Amer. Math. Soc. {\bf 365} (2004) 659--689},
\newblock \href{https://arxiv.org/abs/math/0203250}{\texttt{arXiv:math/0203250}}.

\bibitem{Steph:2005}
K.~ Stephenson, 
\newblock {\em Introduction to circle packing. The theory of discrete analytic
  functions}.
\newblock Cambridge University Press, Cambridge, 2005.

\bibitem{ABS}
V.~E.~Adler, A.~I.~ Bobenko,  and Y.~B.~Suris, 
\newblock Classification of integrable equations on quad-graphs. {T}he
  consistency approach.
\newblock \href{https://doi.org/10.1007/s00220-002-0762-8}{Comm. Math. Phys. {\bf 233} (2003) 513--543},
\newblock \href{https://arxiv.org/abs/nlin/0202024}{\texttt{arXiv:nlin/0202024}}.

\bibitem{Bazhanov:2010kz}
V.~V. Bazhanov and S.~M. Sergeev, 
\newblock A master solution of the quantum
  Yang-Baxter equation and classical discrete integrable equations,
\newblock \href{https://dx.doi.org/10.4310/ATMP.2012.v16.n1.a3}{Adv. Theor. Math. Phys., 
\textbf{16} (2012) 65 -- 95},
\newblock
\href{http://arxiv.org/abs/1006.0651}{{\tt arXiv:1006.0651}}.


\bibitem{Faddeev:1999}
L.~Faddeev, 
\newblock Modular double of a quantum group, 
\newblock in {\em Conf\'{e}rence
  Mosh\'{e} Flato 1999, Vol. I (Dijon)}, vol.~21 of {\em Math. Phys. Stud},
  pp.~149--156.
\newblock Kluwer Acad. Publ., Dordrecht, 2000,
\newblock \href{https://arxiv.org/abs/math/9912078}{\texttt{arXiv:math/9912078}}

\bibitem{Spiridonov-essays}
V.~P. Spiridonov, 
\newblock Essays on the theory of elliptic hypergeometric
  functions,
\newblock \href{http://dx.doi.org/10.1070/RM2008v063n03ABEH004533}{Uspekhi Mat. Nauk
  {\bf 63} (2008) no.~3(381), 3--72},
\newblock \href{https://arxiv.org/abs/0805.3135}{\texttt{arXiv:0805.3135}}.


\bibitem{Skl82}
E.~K. Sklyanin, 
\newblock Some algebraic structures connected with the
  {Y}ang-{B}axter equation, 
\newblock \href{https://doi.org/10.1007/BF01077848}{Func. Anal. Appl. {\bf 16} (1982) no.~4,
  263--270.}




\bibitem{Spiridonov-beta}
V.~P. Spiridonov,
\newblock On the elliptic beta function.
\newblock Uspekhi Mat. Nauk {\bf 56 (1)} (2001) 181--182.
\newblock \href{https://doi.org/10.1070/RM2001v056n01ABEH000374}{English translation: Russ. Math. Surveys {\bf 56} (1) (2001) 185--186}.

\bibitem{Bazhanov:1989nc}
V.~V.~Bazhanov and Y.~G.~Stroganov,
\newblock {Chiral Potts model as a descendant of the six vertex model},
\newblock\href{https://doi.org/10.1007/BF01025851}
{J. Statist. Phys. \textbf{59} (1990) 799--817}.



\bibitem{Au-Yang:2018}
H.~Au-Yang and J.~H.~H.~Perk, 
\newblock Integrable Chiral Potts Model and the 
Odd-Even Problem in Quantum Groups at Roots of Unity. 
\newblock \href{https://arxiv.org/abs/1806.03359}{\texttt{arXiv:1806.03359}}.






\bibitem{BBP90}
R.~J.~Baxter, V.~V.~Bazhanov and J.~H.~H.~Perk,
\newblock Functional relations for transfer matrices of the chiral {P}otts
  model.
\newblock \href{https://doi.org/10.1142/S0217979290000395}{Internat. J. Modern Phys. B {\bf 4} (1990) 803--870.}




\bibitem{Vildanov:2012}
N.~M.~Vildanov,  
\newblock Some extensions of Ramanujan's ${}_1 \psi_1$ summation formula.
\newblock \href{https://arxiv.org/abs/1204.6569}{\texttt{arXiv:1204.6569}}.




\bibitem{Garoufalidis:2017xah}
S.~Garoufalidis and R.~Kashaev, 
\newblock A meromorphic extension of the 3D Index,
\newblock \href{https://doi.org/10.1007/s40687-018-0166-9}{Res Math Sci \textbf{6} (2019) 8},
\newblock \href{https://arxiv.org/abs/1706.08132}{\texttt{arXiv:1706.08132}}.





\bibitem{BKMS}
V. ~V. ~Bazhanov, R. ~M. ~Kashaev, V. ~V. ~Mangazeev and Y. ~G. ~Stroganov,
\newblock ``$({Z}_{N})^{\otimes(n-1)}$ generalization of the chiral
          {P}otts
  model'',
\newblock \href{https://doi.org/cmp/1104202949}{Comm. Math. Phys. {\bf 138} (1991) 393--408}.

\bibitem{Bazhanov:2022wdj}
V.~V.~Bazhanov and S.~M.~Sergeev,
\newblock An Ising-type formulation of the six-vertex model,
\newblock \href{https://doi.org/10.1016/j.nuclphysb.2022.116055}{
Nucl. Phys. B \textbf{986} (2023) 116055},
\newblock \href{https://arxiv.org/abs/2205.10708}{\texttt{arXiv:2205.10708}}.



\bibitem{Faddeev:1993rs}
L.~D.~Faddeev and R.~M.~Kashaev,
\newblock Quantum Dilogarithm,
\newblock \href{htpps://doi.org/10.1142/S0217732394000447}{Mod. Phys. Lett.  A \textbf{9} (1994) 427--434},
\newblock \href{https://arxiv.org/abs/hep-th/9310070}{\texttt{arXiv:hep-th/9310070}}.




\bibitem{Bazhanov:1992jq}
V.~V.~Bazhanov and R.~J.~Baxter,
\newblock Star triangle relation for a three-dimensional model,
\newblock \href{https://doi.org/10.1007/BF01049952}{J. Statist. Phys. \textbf{71} (1993) 839--864},
\newblock \href{https://arxiv.org/abs/hep-th/9212050}{\texttt{arXiv:hep-th/9212050}}.




\bibitem{Bax1}
R.~J.~Baxter, 
\newblock Solvable eight-vertex model on an arbitrary planar lattice.
\newblock Philos. Trans. Roy. Soc. London Ser. A {\bf 289} (1978) 315--346.

\bibitem{Bax2}
R.~J.~Baxter,
\newblock Functional relations for the order parameters of the chiral {P}otts
model.
\newblock \href{https://doi.org/10.1023/A:1023096408679}{J. Statist. Phys. {\bf 91} (1998) 499--524},
\newblock \href{https://arxiv.org/abs/cond-mat/9808120}{\texttt{arXiv:cond-mat/9808120}}.


\bibitem{Str79}
Y.~G.~Stroganov,
\newblock A new calculation method for partition functions in some lattice
  models.
\newblock \href{https://doi.org/10.1016/0375-9601(79)90601-7}{Phys. Lett. A {\bf 74} (1979) 116--118}.

\bibitem{Zam79}
A.~B.~Zamolodchikov, 
\newblock {$Z_{4}$}-symmetric factorized {$S$}-matrix in two space-time
  dimensions.
\newblock \href{https://doi.org/10.1007/BF01221446}{Comm. Math. Phys. {\bf 69} (1979) 165--178}.

\bibitem{Bax82inv}
R.~J.~Baxter, 
\newblock The inversion relation method for some two-dimensional exactly solved
  models in lattice statistics.
\newblock \href{https://doi.org/10.1007/BF01011621}{J. Statist. Phys. {\bf 28} (1982) 1--41}.




\bibitem{Bazhanov:2016ajm}
V.~V.~Bazhanov, A.~P.~Kels and S.~M.~Sergeev,
\newblock
Quasi-classical expansion of the star-triangle relation and integrable  systems on quad-graphs,
\newblock \href{https://doi.org/10.1088/1751-8113/49/46/464001}{
J. Phys. A \textbf{49}, no.46 (2016) 464001},
\newblock \href{https://arxiv.org/abs/1602.07076}{\texttt{arXiv:1602.07076}}.




\bibitem{Chicherin:2017cns}
D.~Chicherin, V.~Kazakov, F.~Loebbert, D.~M\"uller and D.~l.~Zhong,
\newblock Yangian Symmetry for Bi-Scalar Loop Amplitudes,
\newblock \href{https://doi.org/10.1007/JHEP05(2018)003}{
JHEP \textbf{05} (2018) 003},
\newblock \href{https://arxiv.org/abs/1704.01967}{\texttt{arXiv:1704.01967}}.




\bibitem{Kazakov:2023nyu}
V.~Kazakov, F.~Levkovich-Maslyuk and V.~Mishnyakov,
\newblock Integrable Feynman Graphs and Yangian Symmetry on the Loom,
\newblock \href{https://arxiv.org/abs/2304.04654}{\texttt{arXiv:2304.04654}}.




\bibitem{FanWu}
C.~ Fan and F.~ Y.~ Wu, 
\newblock General Lattice Model of Phase Transitions,
\newblock \href{https://doi.org/10.1103/PhysRevB.2.723}{
Phys. Rev. B {\bf 2} (1970) 723-733}.

\bibitem{Felderhoff}
B.~U.~ Felderhof, 
\newblock Diagonalization of the transfer matrix of the free-fermion
model II,
\newblock  \href{https://doi.org/10.1016/0031-8914(73)90330-3}{Physica {\bf 66} (1973) 279–297}.

\bibitem{Bazhanov:1984ji}
V.~V.~Bazhanov and Y.~G.~Stroganov,
\newblock Hidden Symmetry of the Free Fermion Model. 2. Partition Function,
\newblock
\href{https://doi.org/10.1007/BF01017909}{Theor. Math. Phys. \textbf{63} (1985) 519}.

\bibitem{Baxter:1986df}
R.~J.~Baxter,
\newblock Free-fermion, checkerboard and Z-invariant lattice models in
statistical mechanics,
\newblock \href{https://doi.org/10.1098/rspa.1986.0016}{
Proc. Roy. Soc. Lond.  A \textbf{404} (1986) 1-33}

\bibitem{AuYang99}
H.~Au-Yang and J.~H.~H.~Perk, 
\newblock 
The large-$N$ limits of the chiral Potts model, 
\newblock \href{https://doi.org/10.1016/S0378-4371(98)00386-0}{Physica A {\bf 268} (1999) 175–206}, 
\newblock \href{https://arxiv.org/abs/math/9906029}{\texttt{arXiv:math/9906029}}.

\bibitem{Drinfeld:1987}
V.~G.~Drinfel'd, 
\newblock Quantum groups.
\newblock In {Proceedings of the International Congress of Mathematicians,
  Vol. 1, 2 (Berkeley, Calif., 1986)}, pages 798--820, Providence, RI, 1987.
   Amer. Math. Soc.;
\newblock \href{https://doi.org/10.1007/BF01247086}{J. Math. Sci. {\bf 41} (1988) 898–915}.

\bibitem{Jimbo:1986a}
M.~Jimbo,
\newblock ``A {$q$}-analogue of {$U(gl(N+1))$}, {H}ecke algebra, and the
  {Y}ang-{B}axter equation,''
\newblock \href{https://doi.org/10.1007/BF00400222}{Lett. Math. Phys. {\bf 11} (1986) 247--252}.

\bibitem{Faddeev:1987ih}
L.~D.~Faddeev, N.~Y.~Reshetikhin and L.~A.~Takhtajan,
\newblock
Quantization of Lie Groups and Lie Algebras,
\newblock
\href{http://www.mathnet.ru/php/archive.phtml?wshow=paper&jrnid=aa&paperid=7&option_lang=eng}{Alg. Anal. \textbf{1} (1989) 178-206}.

\bibitem{Caselle:2019khe}
M.~Caselle, A.~Nada, M.~Panero and D.~Vadacchino,
\newblock
Conformal field theory and the hot phase of three-dimensional $U(1)$ gauge theory, 
\newblock \href{https://doi.org/10.1007/JHEP05(2019)068}{
JHEP \textbf{05} (2019) 068},
\newblock \href{https://arxiv.org/abs/1903.00491}{\texttt{arXiv:1903.00491}}.





\bibitem{Berezinskii:1971}
V.~L.~Berezinskii, 
\newblock Destruction of long-range order in one-dimensional and two-dimensional systems having a continuous symmetry group I. Classical systems,
\newblock \href{http://jetp.ras.ru/cgi-bin/dn/e_032_03_0493.pdf}{
Sov. Phys. JETP, {\bf 32}(3) (1971) 493–500.}

\bibitem{Kosterlitz:1973}
J.~M.~Kosterlitz and D.~J.~Thouless, 
\newblock Ordering, metastability and phase transitions in two-dimensional systems,
\newblock \href{https://doi.org/10.1088/0022-3719/6/7/010}{
J. Phys. C: Solid State Phys., {\bf 6} (7) (1973) 1181–1203}.

\end{thebibliography}
\end{document}